# Topological edge and corner states in Bi fractals on InSb

R. Canyellas[1,2*], Chen Liu[3*], R. Arouca[4], L. Eek[1], Guanyong Wang[3], Yin Yin[3], Dandan Guan[3,5], Yaoyi Li[3,5], Shiyong Wang[3,5], Hao Zheng[3,5], Canhua Liu[3,5], Jinfeng Jia[3,5#], C. Morais Smith[1]

[1]Institute for Theoretical Physics, Utrecht University, Princetonplein 5, 3584CC Utrecht, The Netherlands
[2]Institute for Molecules and Materials, Radboud University, Heyendaalseweg 135, 6525AJ Nijmegen, The Netherlands
[3]Key Laboratory of Artificial Structures and Quantum Control (Ministry of Education), Shenyang National Laboratory for Materials Science, School of Physics and Astronomy, Shanghai Jiao Tong University, Shanghai 200240, China
[4]Department of Physics and Astronomy, Uppsala University, Uppsala, Sweden
[5]Tsung-Dao Lee Institute, Shanghai Jiao Tong University, Shanghai 200240, China

 * shared first authors
 # corresponding author

(Dated: August 1, 2023)

**Topological materials hosting metallic edges characterized by integer quantized conductivity in an insulating bulk have revolutionized our understanding of transport in matter. The topological protection of these edge states is based on symmetries and dimensionality. However, only integer-dimensional models have been classified, and the interplay of topology and fractals, which may have a non-integer dimension, remained largely unexplored. Quantum fractals have recently been engineered in metamaterials, but up to present no topological states were unveiled in fractals realized in *real* materials. Here, we show theoretically and experimentally that topological edge and corner modes arise in fractals formed upon depositing thin layers of bismuth on an indium antimonide substrate. Scanning tunneling microscopy reveals the appearance of (nearly) zero-energy modes at the corners of Sierpiński triangles, as well as the formation of outer and inner edge modes at higher energies. Unexpectedly, a robust and sharp depleted mode appears at the outer and inner edges of the samples at negative bias voltages. The experimental findings are corroborated by theoretical calculations in the framework of a continuum muffin-tin and a lattice tight-binding model. The stability of the topological features to the introduction of a Rashba spin-orbit coupling and disorder is discussed. This work opens the perspective to novel electronics in real materials at non-integer dimensions with robust and protected topological states.**

Dissipationless transport at the edges of topological insulators holds the promise to revolutionize our technology. Indeed, these topological modes protected by symmetry are insensitive to dopants or disorder, thus preventing scattering and energy losses. The drawback is that the firstly discovered topological state of matter in semiconducting quantum wells, the integer quantum Hall effect[1], required *extremely low temperatures* and *very high magnetic fields*, thus hampering their use in daily life. Further theoretical proposals[2,3] and experimental realization of quantum spin Hall insulators based on the intrinsic spin-orbit coupling eliminated the need of high magnetic fields, although the low-temperature challenge remained[4]. In parallel, the quantum Hall effect has been realized in graphene samples at room temperature, but a very high magnetic field (B = 29T) was required[5]. It was only with the synthesis of bismuthene, a reconstructed honeycomb monolayer of Bi atoms on top of SiC(0001), that the conditions to realize a topological state of matter at room temperatures, without the need of a magnetic field, have been met[6]. This is because the Bi atoms have Z = 83, which leads to a very strong intrinsic spin-orbit coupling, and hence guarantees a robust gap hosting a Z2 topological phase[7–9].

Currently, topological states of matter are being intensively studied, and their classification has been extended to more complex situations, such as out-of-equilibrium (Floquet)[10], non-Hermitian[11], crystalline[12,13], and even interacting systems[14]. Nevertheless, the investigation of topological states at *non-integer dimensions* is still at its infancy. Fractals are known to exhibit a non-integer Hausdorff dimension, and the recent realization of fractals in electronic[15] and photonic[16] quantum-simulator setups has attracted further attention to the topic. Theoretical studies of the integer quantum Hall effect in fractal structures were mostly based on Greens function[17,18] and the Buttiker-Landauer formalism[19,20] to describe transport properties, or on topological markers[17,18,21]. The conditions for the existence and robustness of this topological phase were identified, but no experiments have yet verified their predictions. More recently, higher-order topological (HOT) corner modes were theoretically proposed[22] and experimentally measured in acoustic metamaterials[23,24]. These modes were shown to arise in a 2D Su-Schrieffer-Heeger (SSH) model[23] or in a system[24] corresponding to a generalization of the Benalcazar, Bernevig and Hughes (BBH) model[25,26], which requires the implementation of quadrupolar order. However, until now there were no theoretical or experimental investigations of the *quantum spin Hall effect* in fractals. Moreover, all the above-mentioned experiments were realized in engineered materials, and the observation of topological states of matter in *real condensed-matter fractals* remained a challenge. The reason is because planar fractals optimize the perimeter to area ratio, requiring the covalent interaction on the bonds between atoms to be more relevant than the van der Waals interaction[27–29], which is volumetric, and this is rarely met.

Recently, Bi monolayers were deposited on an InSb(111)B substrate, and upon varying the annealing temperature and time, Sierpiński triangles were spontaneously formed[28]. Later, similar structures were synthesized and studied using reflective high-energy electron diffraction and core-level photoelectron spectroscopy[30]. These single-element fractals self-assembled on a semiconducting surface[28] provide an ideal platform for the investigation of topological states at non-integer dimensions, since Bi is a heavy element, and hence a candidate to exhibit a robust quantum spin Hall effect. In this work, we show theoretically and experimentally that topological edge and corner modes arise in fractal Bi monolayers on InSb substrate. This is the first observation of topological states of matter in real condensed-matter fractals, which lays a foundation for the research of topological states at non-integer dimensions, and the preparation of quantum devices.

In our experiments, we deposit Bi atoms on the prepared InSb(111)B substrate surface (see Supplementary Materials（SM）for details) at 423K. After deposition, we cool the system to room temperature for one hour. We measure the STM images, dI/dV spectra, and local density of states (LDOS) maps at liquid-helium temperature, of about 4 K. The STM images in Figures **1a-1c** show one monolayer (1-ML) Bi films in the shape of Sierpiński -triangles formed on the InSb(111)B substrate surface. In Figure **1d**, the different film height levels are depicted and labeled: 0, $S_1$, 2, and $S_3$ from lower to higher (see SM for details). Level 0 is assumed to be a wetting layer composed of Bi, Sb, and In atoms, level $S_1$ and level $S_3$ correspond to a Sierpiński triangle-like 1-ML Bi on the wetting layer and on level 2, respectively. Level 2 is a 2-ML Bi with (2×2)-reconstructed or disordered[28]. Next, we study the distribution of electronic states in this sample. Figure **1e** shows the dI/dV spectra measured on the InSb substrate (sub) and on the wetting layer (level 0). Near the Fermi surface, the semiconducting energy gap of the InSb substrate and of the wetting layer are about 0.3eV and 0.1eV, respectively, which is beneficial to the study of topological states in this system. Figure **1f** shows the dI/dV spectra measured on different positions of the Sierpiński triangle Bi monolayer along the edge represented in Figure **1c**, where the red arrow denotes the measurement direction, and the blue dots are the measurement positions. There are many characteristic peaks at different bias energies, and the semiconducting gap near the Fermi surface has basically disappeared. These spectra exhibit similar peaks at some bias energies, but their intensity is different, especially near the Fermi surface. The peaks suggest that there are novel electronic properties in the sample, which we now discuss in combination with theoretical calculations. Especially notable is a V-shaped zero-energy peak at the corners of the sample (cyan), which hints at a possible topological state[31].

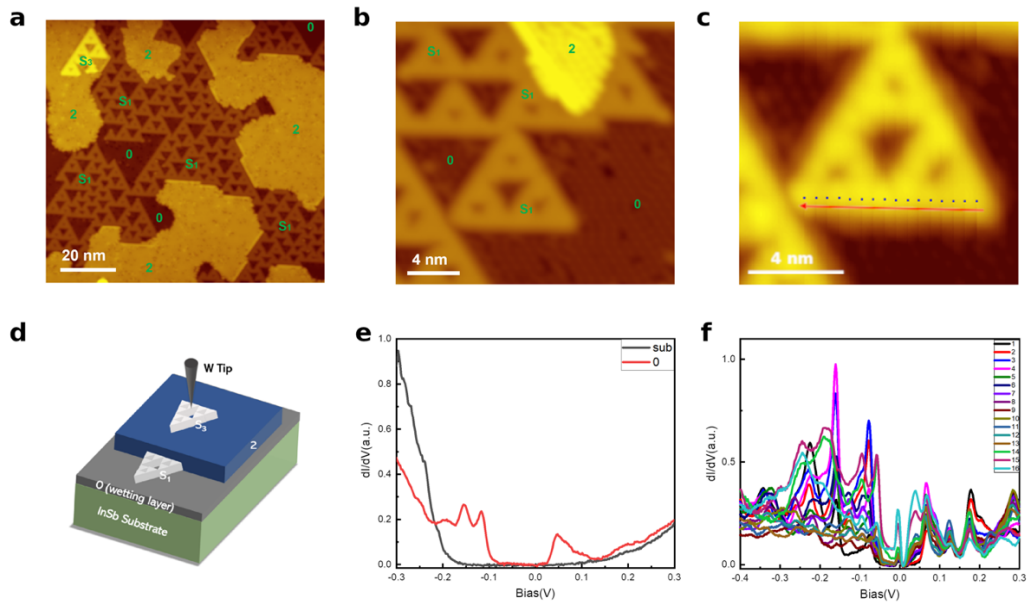

**Figure 1: Fractal Bi monolayers on InSb substrate a, b, c** STM images of Sierpiński -triangle Bi monolayers formed on InSb(111)B surface. **d** Schematic for the sample, illustrating the different film-height levels. **e** dI/dV spectra measured on the InSb substrate surface (sub) and on the wetting layer (level 0), respectively. **f** dI/dV spectra measured on different positions of the Sierpiński -triangle Bi monolayer along the blue dots shown in **c**, where the red arrow indicates the measurement direction, and spectra numbered from 1 to 16 are taken on the blue dots.

## Results

### 1.  Theoretical calculations of the LDOS

Our theoretical calculations based on a variation of the muffin-tin model reveal that indeed, corner modes are expected near zero energy, and robust edge states should appear at the outer and inner edges at different energies. These calculations implement the spin-orbit coupling within a continuum model, which is a Schrödinger equation describing the electrons confined by the fractal potential. Since the intrinsic spin-orbit term is proportional to a derivative of the confining potential, a model using sharp barriers is inadequate, and for this reason we implement more realistic smooth Gaussian potentials, see Methods and Supplementary Materials (SM) for more details.

A comparison of the local density of states (LDOS) evaluated for the purely kinetic term and for the full model, including kinetic and intrinsic spin-orbit coupling, is presented in Figs. **2a-2e** (pink background) and **2f-2j** (purple background), respectively. The LDOS maps are shown at selected values of energies, for which some lattice sites exhibit a characteristic feature, like same energy, a peak, or a dip. The lattice sites are colored according to their connectivity: red sites have only one neighbor, blue sites have three neighbors, and green sites have two neighbors, see inset of Fig. **2k**. The results clearly reveal that the addition of the intrinsic spin-orbit coupling to the Schrödinger equation (purple background) leads to the appearance of zero-energy modes at the outer corners (see LDOS in Fig. **2h**), and to edge modes at the outer and inner boundaries of the Sierpiński *gasket* (see Figs. **2i** and **2j**, respectively). These modes could have a topological origin.

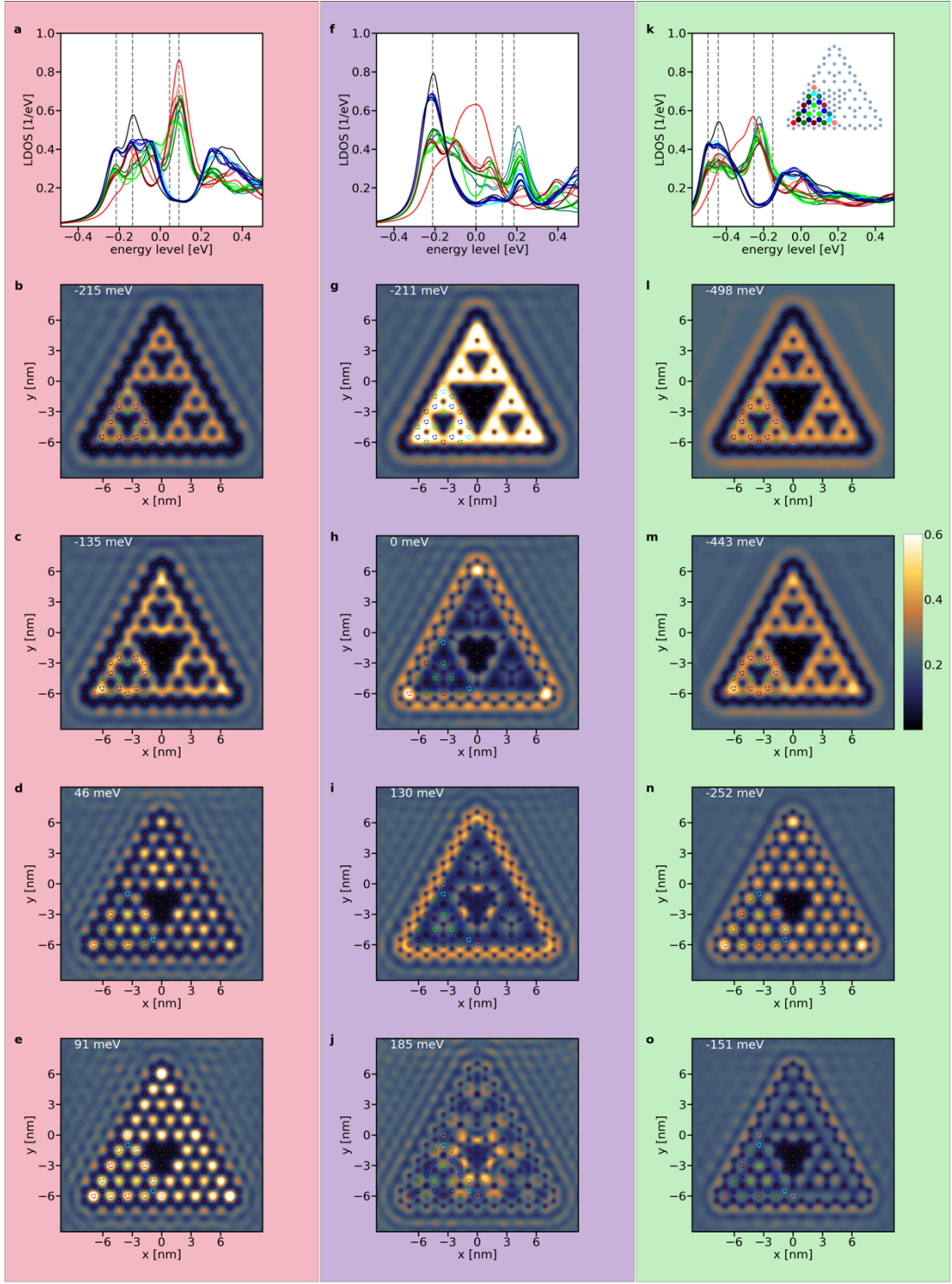

**Figure 2: LDOS maps for the 3rd generation Sierpiński triangle obtained theoretically using the muffin-tin method with a smooth Gaussian potential.** The calculations shown in the first column (pink) include only the kinetic term, while the second column (purple) also includes the intrinsic spin-orbit, and the third (green) includes all the three terms, kinetic, intrinsic, and Rashba spin-orbit coupling. The number of grid points in both axes are $n_x = n_y = 200$, the number of waves is 1500, the full width at half maximum (FWHM) of the Gaussian potential is d = 0.62nm, and the potential height u=0.9eV. The effective electron mass $m_{eff} = 0.42$, in units of the bare electron mass, the lattice parameter $a_0$ =1nm, and the intrinsic and Rashba spin-orbit coupling are $\lambda_{ISOC} = 10^6$ and $\lambda_{Rsh} = 10^9$, respectively. These parameters have been set such that the Rashba and intrinsic spin-orbit terms have the same order of magnitude.

In addition, several features are worth noticing. Firstly, at E = 130meV (Fig. **2i**), outer edge modes coexist with a high LDOS in the *center* of the inner-triangle edges. Secondly, there is intensity missing at the corners of the edge state localized at the larger inner triangle at E = 185meV (Fig. **2j**). Both features are reminiscent of the ones theoretically predicted [22] and experimentally observed[23,24] for a Sierpiński *carpet* realized with acoustic metamaterials. The main difference is that in Refs. 22-24 a HOT Sierpiński carpet was investigated, whereas here we have a different crystalline symmetry, $C_3$ instead of $C_4$, and the topological properties are driven by a spin-orbit coupling, instead of a Wilson mass[22], a dimerization[23], or a magnetic flux[24]. Moreover, in a true HOT insulator[24] the zero-energy corner modes appear in a gapped edge state, while here the edge states have a lower, but non-zero LDOS. Zero-energy corner modes generated by destructive interference and protected by latent symmetry, like the ones observed in a breathing Kagome[32,33] or in a diamond necklace chain[34], were recently shown to emerge in an s-orbital tight-binding description of a Sierpiński triangle with honeycomb symmetry[35]. These features are similar to the ones describing boundary obstructed insulators[33,36,37], which are predicted by Ref. [36] to be present in bismuthene. However, the problem considered here is more intricate because the bulk parts of the Sierpiński have a triangular structure instead of honeycomb, and several orbitals, s, $p_x$, and $p_y$ are involved (see Methods for details).

A direct way to verify whether these edge and corner modes are topological or not is to investigate their resilience to the introduction of a Rashba spin-orbit coupling. In the quantum spin Hall phase, the edge modes are destroyed by a Rashba coupling[2,38], which may close the bulk bandgap and bring the system into a trivial state. In Figs. **2k-2o** (green background), one can observe the effect of a Rashba spin-orbit coupling on the LDOS. The edge and corner modes are barely visible in the LDOS maps, and the spectrum (Fig. **2k**) became very similar to the purely kinetic case, in which the LDOS at the green and red sites exhibit a peak at the same energy. Since the inclusion of the Rashba term has shifted the peaks, we present here the LDOS maps at the most prominent peak values. Its effect for precisely the same values of energy selected in Fig. **2f** is presented in the SM for completeness.

## 2. Comparison between theoretical and experimental LDOS

More evidence of the adequacy of the theoretical description to capture the features measured experimentally is provided by comparing the LDOS maps at several bias voltages. We have selected the most relevant mappings and adopted a similar color code for theory and experiments. In addition, we complemented the muffin-tin theory with tight-binding calculations, accounting for several orbitals (see Methods and SM for details). Fig. **3a** (top) shows the STM image of a second generation Sierpiński triangle, on which the STM spectra will be taken, as well as a sketch of the theoretical setups. The muffin-tin calculations are shown for a third generation, since more features are visible at higher generations. Calculations for lower generations are shown in the SM for completeness. The electronic sites are denoted by red, blue, and green dots, and the potential barriers are represented yellow dots surrounded by a purple circle. The tight-binding calculations are performed for the geometry reconstructed from the experimental observations. Figs. **3b-3g** show six sets of three figures, with the experimental measurements on the top and the corresponding theoretical simulations, muffin tin at the middle and tight-binding at the bottom. The first two experimental images for the lowest values of the bias voltage, at V = −507mV and V = −266mV, exhibit a good agreement with the theoretical LDOS depicted below them. Both show a high intensity inside the Sierpiński triangle, characterizing a bulk phase. A very interesting and unexpected feature is that the fractal boundary is perfectly defined by a cyan line at V = −507mV, denoting a very low density of states. These sharp and well-defined depleted lines appear for a broad range of voltages in all samples, independently of disorder or any other factor, like the generation of the fractal. They seem to be a robust and recurrent feature in our measurements (see Fig.4b and Fig.15 in the SM for further experimental observations). The next experimental image, at V = −104mV, exhibits four blobs along the edges. Although the theoretical muffin-tin image is not exactly the same, the tight-binding one is closer to the experimental map. It is also possible that there is a missing pink dot at the right bottom of the experimental map due to disorder, and that two blobs have merged at the left bottom corner. This would correspond to the five pink blobs along the Sierpiński triangle edges observed in the muffin-tin image. The next experimental image at V = 16mV also exhibits a good agreement with theory. Corner modes are visible, but there seems to be a missing corner mode at the bottom right of the experimental image, most probably due to disorder. We will come back to this point later, when we investigate the effects of different types of boundaries and disorder on the LDOS.

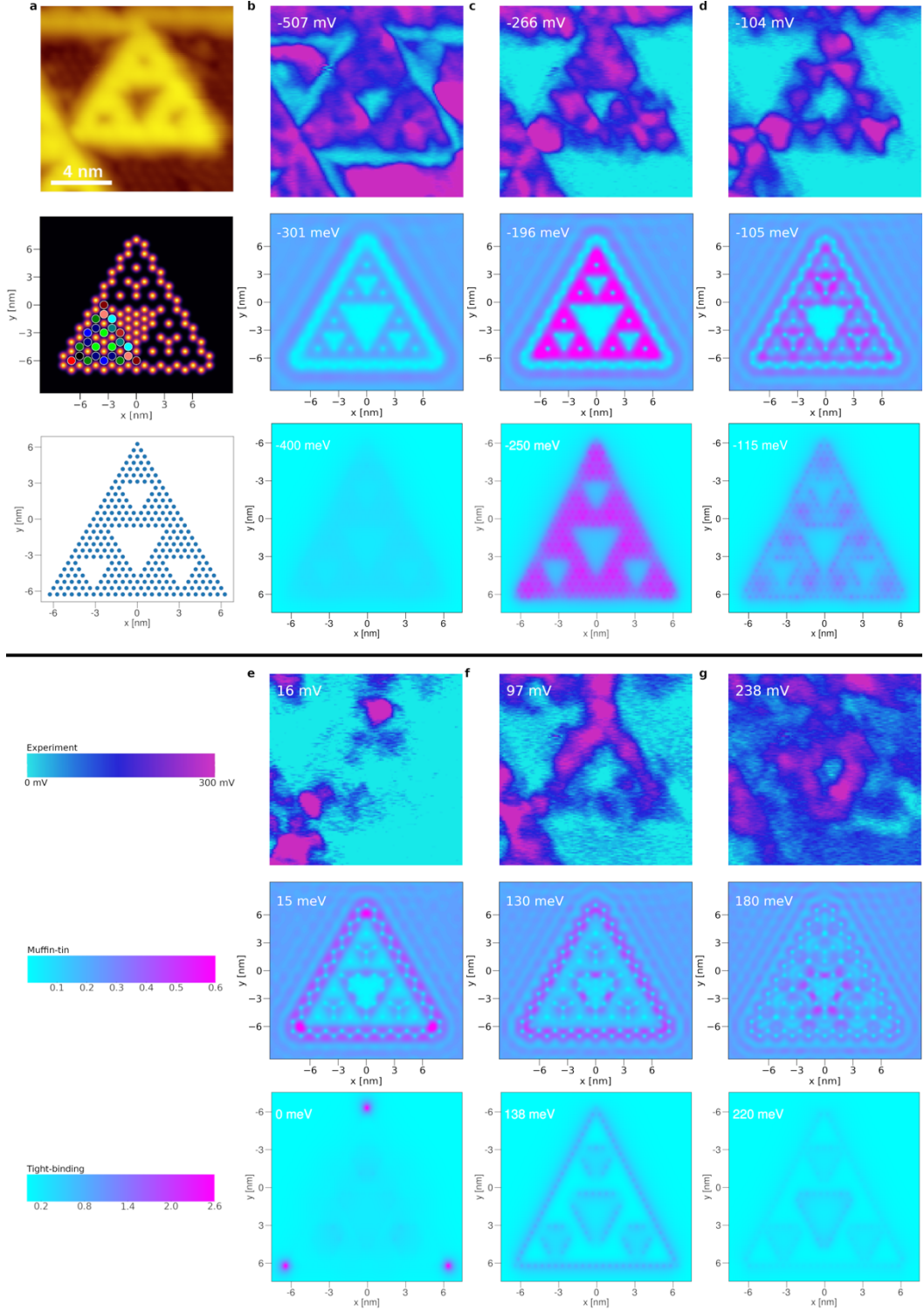

**Figure 3: LDOS for several values of the bias voltage. a** Topographic image of the fractal measured with the STM (top); sketch of the muffin-tin model (middle) and of the tight-binding lattice (bottom). Yellow dots surrounded by purple lines represent the scatterers. **b-g** Experimental LDOS (top); LDOS calculated within the muffin-tin (middle) and within the tight-binding model (bottom). Experimental and theoretical color bars are shown below the sketches. The theoretical muffin-tin simulations are performed in a third generation G=3 Sierpiński triangle. The number of grid points in both axes are $n_x = n_y = 200$, the number of waves is 1500. The full width at half maximum (FWHM) of the Gaussian potential is d = 0.62nm and the potential height u=0.9eV. The effective electron mass $m_{eff} = 0.42$, the lattice parameter $a_0 = 1$nm, and the intrinsic spin-orbit parameter $\lambda_{ISOC} = 10^6$.

Moving onto the next set of figures, we notice that the LDOS measured experimentally at V = 97mV exhibits an edge state (pink) along the outer Sierpiński triangle boundary, while the interior remains cyan, corresponding to a low LDOS. The theoretical maps also show images where the boundary surrounding the Sierpiński triangle is pink, which is suggestive of an outer edge state. The sharp dots expected to appear concomitantly at the center of the inner edges, observed in the HOT phase of acoustic metamaterials[23,24] are not visible in our experiments, possibly due to disorder. From the outer perimeter, the LDOS gradually goes to the inner perimeter upon increasing the energy, as observed for V = 238mV. Notice that the pink triangle (high LDOS) has inverted from upward to downward. At the corresponding theoretical image, the pink region is mainly localized around the large inner hole and the corners of the inner triangle are depleted, as expected theoretically. Indeed, these seem to be reminiscent of the type II trimer phases predicted in Ref. 23, in which the corner of the trimer has zero intensity. Finally, we note that the tight-binding results are more faded because we kept the same color code for all images. If individual variations are allowed, the features become more prominent (see SM).

We discussed in detail the agreement of our calculations with the most prominent results observed experimentally, but this holds also for other energies, below and above the ones shown in Fig. 3. We refer the reader to the SM for a full comparison at all energies. Notice, however, that although the sequence of theoretical images corresponds to the features observed experimentally, we do not have agreement for the precise values of energy. This will be improved below, after considering different types of boundaries and disorder.

### 3. Effects of disorder

Next, we investigate the effect of disorder within the muffin-tin model, which is the closest to the experimental realization. Inspection of the topographic images obtained experimentally show that the structures are not perfect. Indeed, it can be observed in Fig. 3 that the LDOS does not have a $C_3$ rotational symmetry, as one would expect from a perfectly clean Sierpiński gasket fractal. There is clearly a geometric disorder, but there could also be chemical structural disorder, which would generate a potential difference in the corresponding region, or defects. To simulate these cases, we created several disordered structures. Firstly, we investigated *geometric* disorder, in which the sites of the Sierpiński triangle are not perfectly positioned. Then, we engineered *potential* disorder by randomly varying the height of the potential barriers, and *displacement* disorder, in which the center of the Gaussian potentials is away from the honeycomb underlying grid. Finally, for a *straight potential boundary*, we included *rotation and disorder on the box boundary* (position of points) to show that a quantitative agreement is possible between theory and experiments.

*Geometric disorder*. Since the experimental samples are geometrically disordered, we developed an algorithm that generates pseudo-random samples that mimic the experimental results. The building blocks are the anti-lattice of a generation G and the central equilateral triangle corresponding to the central hole of a generation G′. By combining these two and performing translations along the diagonals of the anti-lattice, we generate potential landscapes such as the one shown in Fig. **4a**. More details are provided in the Methods. This type of disorder does not have a strong influence, and the LDOS profiles are the same inside each Sierpiński triangle (see Figs. **4b-e**, where a bulk phase, corner modes, outer edges, and inner edges are observed as the energy is increased). The main effect is that regions closer to many scatterers exhibit a more intense LDOS. In general, parts of concatenated triangles do not interfere with each other. Disorder might kill one corner mode, e.g., but the rest remains unperturbed.

*Potential disorder*. Another type of disorder consists of random potential barriers, as shown in Fig. **4f**. For each scatterer, we chose the height of the Gaussian potential from a uniform distribution. The height of each scatterer is represented by a color code in Fig. **4f**, with light colors indicating higher potentials. A careful observation of the yellow dots representing the barriers reveals that some are brighter (higher) than others. The results show that the LDOS behavior is still similar (see Fig. **4g**), but with a stronger asymmetry (see Figs. **4h-j**). Upon increasing the range of random increments, but keeping its mean value fixed at u=0.9 eV, we observe that the phases remain qualitatively robust. When the height has an error of about 10% of its original value, some asymmetry appears and phases that were initially higher in energy begin to form earlier in some parts of the Sierpiński triangle (see Fig. **4j**, in which intensities around the smallest triangle appear concomitantly with the outer edge mode). A connection of this type of behavior with the experimental observations is discussed in the SM.

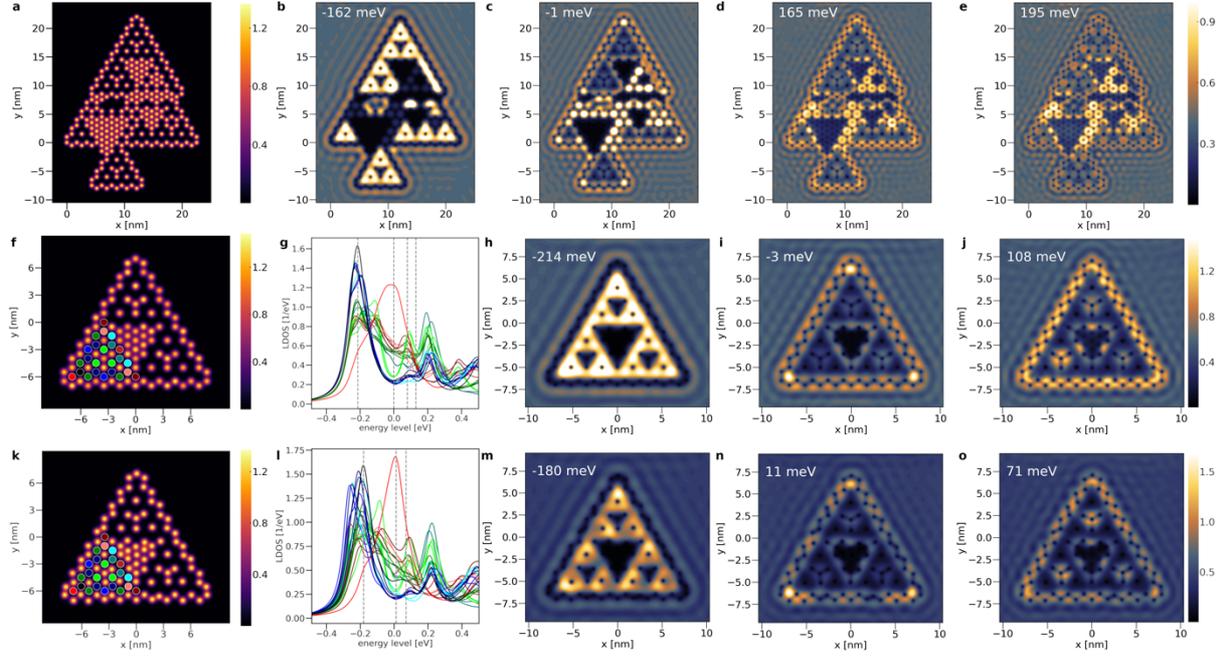

**Fig. 4: Effects of disorder. a-e** Geometric disorder, **f-j** potential disorder, **k-o** displacement disorder. **a,f**, and **k** show the type of disorder studied, and the other plots are the LDOS in presence of the specific disorder. All different types of disorder have been simulated using the same parameters as before; a number of grid points $n_x = n_y = 200$, an effective electron mass $m_{eff} = 0.42$, a lattice parameter $a_0 = 1$nm, an intrinsic spin-orbit parameter $\lambda_{ISOC} = 10^6$, a potential height u=0.9 eV, and FWHM of the Gaussian potential d = 0.62nm. For geometric disorder, we used 1500 waves, but for the other two types of disorder only 750 waves. In the potential disorder, we introduced an error to the potential height taken from a uniform distribution between [-0.1u,0.1u]. For the position disorder, the scatterers coordinates are modified using a uniform distribution between [-0.1$a_0$,0.1$a_0$] for both axes.

***Displacement disorder***. The next type of disorder that we consider consists of moving the center of the Gaussian potentials away from the underlying honeycomb grid, see Fig. **4k**. Here, we notice that all yellow dots representing the barriers have the same intensity, but their position is misplaced. The obtained LDOS is shown in Figs. **4m-4o**. Again, we observe the same phases as before, but disordered. The results for this type of disorder show that as the displacement of the scatterer increases, the obtained LDOS becomes more asymmetric (see SM). However, it is not necessary to introduce a large deviation to already observe its effect, which is a consequence of the interference nature of the LDOS patterns. This type of disorder is highly detrimental to the observed phases.

***Straight potential confinement and disorder on box boundary***. Finally, we consider the same kind of problem, but now the boundary of the Sierpiński triangle is given by a straight line, and not composed of small barriers at the atomic length. In addition, we introduce an asymmetry (rotation) in the way how we position the Sierpiński triangle within the square box to identify further sources of disorder in the quasi-particle interference pattern. Theoretical results of the LDOS for the second generation including only the kinetic term (pink background), kinetic and intrinsic spin-orbit coupling (purple), and all of them plus Rashba (green background) are shown in the SM. Here, we focus on a comparison between the results from the muffin-tin theory and experiments, see Fig. 5. We find that in the presence of an intrinsic spin-orbit coupling and this type of disorder, the right-bottom corner mode disappears (E = 0meV), similarly to what is observed experimentally at V = 16mV. This corroborates our interpretation of disorder as a possible source for the differences observed between theory and experiments. In addition, the LDOS around E = -106meV now shows four blobs along the triangle sides, in agreement with the experiments at V = -104mV, instead of the five obtained in Fig. 3 for the muffin-tin potential. Finally, we would like to emphasize that although we selected some specific values of energy to depict the main features, they remain in a reasonably large range of parameters, thus making the agreement between theory and experiment quantitative.

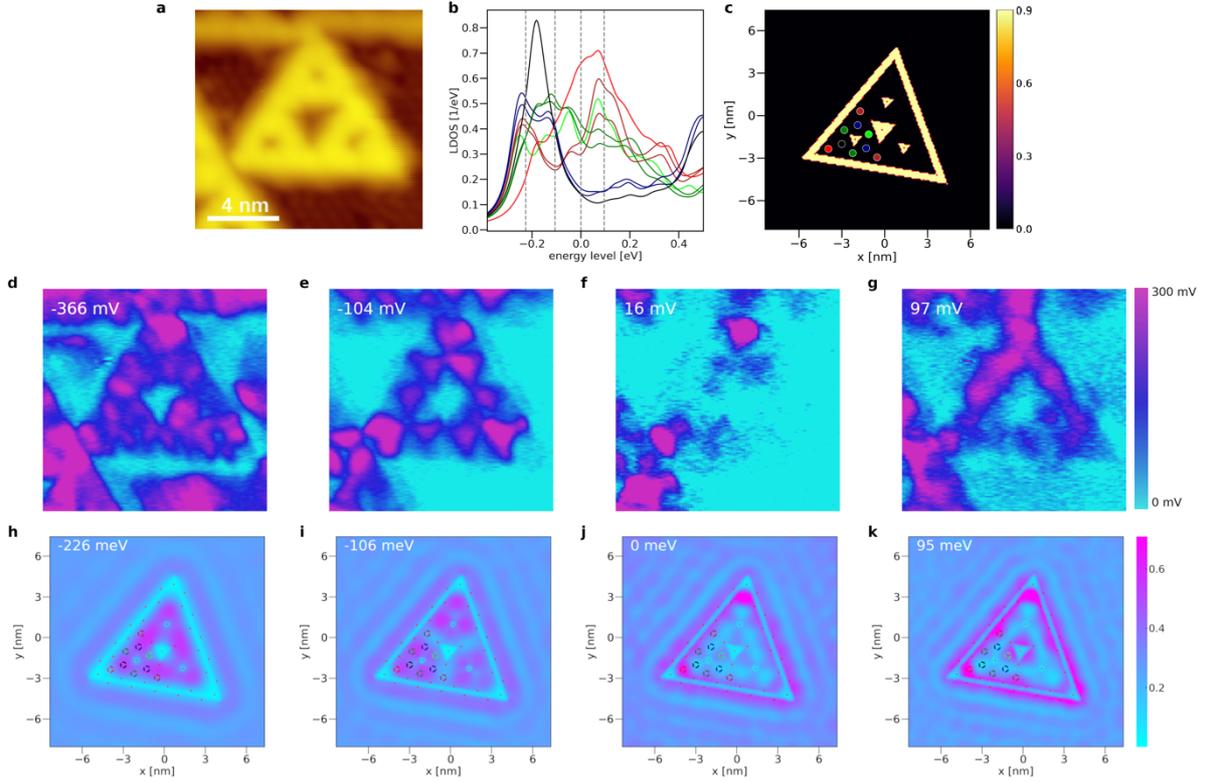

**Fig. 5: Comparison between theoretical and experimental LDOS calculated for a Sierpiński gasket with a straight potential barrier, which is asymmetrically positioned inside a square box.** The number of grid points in both axes are $n_x = n_y = 150$, the number of waves is 750, the width of the potential wall is d = 0.62nm, the rotation angle is θ=10°, and the potential height u=0.9eV. The electron effective mass $m_{eff}$ =0.42, the lattice parameter $a_0$ =1nm, and the intrinsic spin-orbit coupling is $λ_{ISOC} = 10^6$.

## Conclusions and Outlook

Here, we present theoretical and experimental studies of Sierpiński gasket fractals formed spontaneously upon growing Bi thin layers on InSb substrates under very specific conditions (temperature and fluence), see Fig. 1. Since bismuth has a very large spin-orbit coupling, one could expect the realization of the quantum spin Hall effect in these fractal samples, thus providing the first experimental observation of topological states at non-integer dimensions in *real materials*.

To explain and corroborate the experimental findings, we investigate the effect of intrinsic spin-orbit coupling in a Sierpiński triangle by performing quantum simulations in the framework of two different theoretical models, namely a continuum muffin-tin with a smooth Gaussian potential, and a tight-binding description accounting for several orbitals. When the spin-orbit coupling term is of the order of the kinetic term, new phases such as corner modes and edge states are generated. Sharp zero-energy corner modes appear at the vertex of the larger triangle, while edge modes arise at the outer and inner boundaries. These simulations allow for a good qualitative understanding of the sequence of LDOS measured experimentally upon varying the voltage. However, to obtain a quantitative description, the disorder of the fractal structures must be considered. Upon including geometric disorder, we could better simulate the topographic images obtained with the STM. These results show that whenever different Sierpiński triangles touch each other, the LDOS of the whole sample is shifted towards that point. Therefore, if there is a region that has an overall increase of the potential in comparison to the neighbor regions, the LDOS is more pronounced there. We also studied a straight-boundary confining potential. By implementing these configurations and rotating the fractal by ten degrees, we obtained the same two corner modes as observed experimentally, with the right-bottom corner missing. In addition, the edge estates were also shifted, thus supporting the argument of disorder in the real samples.

To verify whether the corner and edge modes have a topological origin, calculations with a Rashba spin-orbit coupling have been performed. As expected, a strong Rashba coupling leads to the disappearance of the corner and edge states. The LDOS then looks more like the one obtained when only the kinetic contribution is present. The disappearance of the edge modes in the presence of Rashba spin-orbit coupling and their resilience to disorder confirms their topological origin. The theoretical investigation of several types of disorder indicates that in the experiments, geometric and potential disorder might be present, but probably there is no displacement disorder because those would be more detrimental and could ruin the topological phases. Therefore, our studies provide not only the first experimental observation of a quantum spin Hall effect at fractional dimensions in a real material but also a consistent explanation of the measurements and robustness of the phases. Notable is the occurrence of a zero-energy (outer) corner mode in the absence of dimerization or magnetic flux, which are known to lead to a HOT insulator. These corner modes appear to be quite resilient to certain types of disorder but could be damaged if the disorder produces destructive interference, as shown by rotating the sample in the theoretical calculations. Finally, a novel feature was observed theoretically and experimentally: extremely sharp and well-defined *depleted edge modes* (cyan lines), with a very low LDOS, were shown to arise at negative voltages, for a very broad range of energies. This hints to a novel type of edge states, not reported to date.

The implications of our findings are multifold. On the one hand, they show that robust topological corner states are available in real materials at fractional dimensions, which could be used in technological devices. The existence of fractionally charged excitations in the fractional quantum Hall effect has fascinated researchers for years, especially because of their potential as qubits for quantum computers. In acoustic fractals, a fractional corner mode was recently shown to emerge from the non-integer dimension of the designed geometry[24]. This indicates that new routes might be available to the generation of edge and corner states of topological origin with non-trivial behavior. Our work has answered many of the open questions in the field of topology at non-integer dimensions, but many more remain to be investigated. Since fractals are pervasive in Nature, the robust topological states found here might have great applications in future.

## METHODS

### Theory 1: Muffin-tin method

To simulate the spontaneously formed Sierpiński fractals, we solve the time-independent Schrödinger equation in the presence of both intrinsic and Rashba spin-orbit coupling

$$H\psi = \left( -\frac{\hbar^2}{2m_e^*}\nabla^2 - i\lambda_{ISOC}\cdot\frac{\hbar^2}{(2m_e^*c)^2}\left((\nabla V)\times\nabla\right)\cdot\sigma - i\lambda_{Rsh}\cdot\frac{\hbar^2}{a_0 m_e^*}(\nabla\times\sigma)\cdot\hat{z} + V \right)\psi, \quad (1)$$

in a square grid $2n_x n_y \times 2n_x n_y$, which accounts for the $x$ and $y$ coordinates and for spins up and down. $\psi$ is a vector in the grid, with each of its components $\psi_s(r)$ describing the wavefunction at position $r = (x, y)$ and spin $s$. The Laplacian $\nabla^2$ and the first derivative $\nabla$ are defined in the grid by using the second-order centred finite-element definition with second-order precision in the grid parameter, with periodic boundary conditions[39]. Here, $\sigma$ denotes the vector of Pauli matrices, the intrinsic and Rashba spin-orbit coupling parameters are represented by $\lambda_{ISOC}$ and $\lambda_{Rsh}$, respectively, and the effective electron mass $m_e^* = 0.42\,m_{el}$, with $m_{el}$ the bare electron mass. The shape of the lattice is determined by the potential $V = \mathrm{diag}(v(\boldsymbol{r}))$, which is a diagonal matrix in the grid with elements proportional to $v(\boldsymbol{r})$. We use *repulsive* potentials to design the anti-lattice of the Sierpiński triangle by selecting positions $\mathbf{R}_i$ where the scatterers are centred and choose a form of the potential around them as shown in the inset of Fig. 2k. In the main text, we use a Gaussian potential

$$v_G(\boldsymbol{r}) = \sum_i \frac{u}{\Delta\sqrt{2\pi}} e^{-\frac{(r-R_i)^2}{2\Delta^2}}, \qquad (2)$$

with potential height $u$ and variance $\Delta$. In simulations, we fix the full width at half maximum (FWHM) $d = 0.62$nm, and compute the corresponding $\Delta = (d/2)\sqrt{2\ln(2)}$. It is necessary to use Gaussian potentials because the Gaussian is a smooth function everywhere and, therefore, prevents spurious divergences of the spin-orbit coupling in regions with step-like potentials. In the SM, we also show results for the muffin-tin potential

$$v_{MT}(\boldsymbol{r}) = \sum_i u\theta(d - |\boldsymbol{r} - \boldsymbol{R}_i|), \qquad (3)$$

where $\theta$ is the Heaviside step function and $d$ describes the diameter of the potential barrier. This potential is more commonly used, but its derivatives are proportional to Dirac delta functions, and therefore it is not adequate in the presence of intrinsic spin-orbit coupling.

**Local density of states**

By solving Eq. (1), we obtain the wavefunction for the different energies $\epsilon$. The LDOS is then promptly calculated using

$$LDOS(E) = \frac{1}{\pi}\sum_\epsilon |\psi_\epsilon|^2 \frac{b}{(\epsilon - E)^2 + (b)^2}, \qquad (4)$$

where $b = 0.04$ eV represents the broadening of the peak at energy $\epsilon$. The LDOS was found to be very sensitive to the method used. For example, enough free space needs to be added around the scatterers to provide a more reliable solution. Hence, the shape and size of the box containing the fractal is also relevant.

**Theory 2: Tight-binding description of the lattice**

Following the approach of Ref. [40], we present an effective tight-binding description of the Bi Sierpiński fractals. Thin layers of Bi are known to form honeycomb [6,41] or triangular [28,42,43] lattice structures, depending on the underlying substrate and deposition method. In Ref. [28], density functional theory (DFT) calculations were performed to explain the structure observed in STM measurements on the Sierpiński fractals investigated here. Based on these results, we consider a triangular lattice in the shape of a Sierpiński fractal, with 'bulky' regions inside the smaller triangles, as shown in Fig. 3a of the main text.

We use a phenomenological minimal model for Bi consisting of three orbitals: $s$, $p_x$ and $p_y$ [40,41,43]. Intrinsic spin-orbit coupling is incorporated through an on-site interaction $\lambda$ between $p_x$ and $p_y$ orbitals. The resulting three-band Hamiltonian governing the system is given by

$$H = H_0 + H_\lambda,$$

$$H_0 = \sum_{i,\alpha} \varepsilon_\alpha c_{\alpha,i}^\dagger c_{\alpha,i} + \sum_{\langle i,j \rangle, \alpha, \beta} t_{\alpha,\beta,i,j} c_{\alpha,i}^\dagger c_{\beta,j},$$

$$H_\lambda = -i\lambda \sum_i \left( c_{p_x,i}^\dagger c_{p_y,i} - c_{p_y,i}^\dagger c_{p_x,i} \right) \sigma_z,$$

where $\varepsilon_\alpha$ is the on-site energy per orbital, $\alpha \in \{s, p_x, p_y\}$, and $t_{\alpha,\beta,i,j}$ are the Slater-Koster hopping parameters between nearest-neighbor sites $i$ and $j$. These parameters are given in Table 1, where they are expressed in terms of the directional cosines $\ell$ and $m$ betweens sites $i$ and $j$, in the $x$ and $y$ directions, respectively. The parameters $t_{ss\sigma}$, $t_{sp\sigma}$, $t_{pp\sigma}$, and $t_{pp\pi}$ represent the hopping parameters between $\sigma$- and $\pi$-hybridized $s$- and $p$-orbitals. The intrinsic spin-orbit term describes an on-site coupling between $p_x$ and $p_y$ orbitals, with strength $\lambda$. This type of spin-orbit coupling is usually much stronger than the $s$-type coupling because of its on-site character, instead of next-nearest neighbor [44].

| | $s$ | $p_x$ | $p_y$ |
|---|---|---|---|
| $s$ | $t_{ss\sigma}$ | $\ell\, t_{sp\sigma}$ | $m\, t_{sp\sigma}$ |
| $p_x$ | $-\ell\, t_{sp\sigma}$ | $\ell^2\, t_{pp\sigma} + (1-\ell^2)t_{pp\pi}$ | $\ell m\big(t_{pp\sigma} - t_{pp\pi}\big)$ |
| $p_y$ | $-m\, t_{sp\sigma}$ | $\ell m\big(t_{pp\sigma} - t_{pp\pi}\big)$ | $m^2\, t_{pp\sigma} + (1-m^2)t_{pp\pi}$ |

**Table 1. The Slater-Koster parameters.**

The parameters were chosen in a way to provide the best fit to the experimental observations. This resulted in $t_{ss\sigma} = 0.072\text{eV}$, $t_{sp\sigma} = -0.1050\text{eV}$, $t_{pp\sigma} = -0.174\text{eV}$, $t_{pp\pi} = 0.0421\text{eV}$, $\varepsilon_s = 0\text{eV}$, $\varepsilon_{p_x} = \varepsilon_{p_y} = 0.0366$ eV, $\lambda = 0.1\text{eV}$.

The LDOS is calculated according to Eq. 4 using the spectrum and eigenstates obtained through numerical diagonalization of the tight-binding Hamiltonian. The tight-binding LDOS is then plotted in a basis of hydrogenic orbitals, such that it can be represented in a continuous manner instead of the discrete tight-binding solutions.

**Theoretical description of disorder**

The different types of disorder shown in Fig. 4 correspond to disorder in the parameters of the simulation. For potential disorder, the values of the barrier height $u$ are randomly selected from a uniform distribution centred on the uniform value $u = 0.9\text{eV}$, while for position disorder we do the same but for $\boldsymbol{R}_i$.

For geometric disorder, we simulated disordered samples by building pseudo-random implementations. The building blocks of these configurations are Sierpiński triangles of generation G ≤ 3, and central holes of generation 2 ≤ G ≤ 4 (the inverted equilateral triangles of scatterers). Following these constrains, we start with the random generation G, then at most N ≤ 3 translations equal to the side of the triangle are applied to the scatterers (or not), in one of the six different directions along the sides. We draw an inverted triangle at the maximum coordinate of either $x$ or $y$, and then another Sierpiński triangle at one of the corners of these last inverted triangles. We repeat this process starting again at the origin.

We also investigated another type of confinement. Instead of defining an anti-lattice and a potential centred there, as in the muffin-tin or in the Gaussian, we constructed a straight boundary with a width that has a constant potential. The inner inverted equilateral triangles have a similar straight constant potential everywhere. We observe that by rotating the fractal inside its confining box leads to disorder in the quasi-particle interference pattern and may lead to the disappearance of some of the corner modes.

## METHODS REFERENCES

**Data availability.** All data supporting the findings of this study is available from the corresponding author upon reasonable request. Source data are provided with this paper.

**Code availability.** The numerical codes used for solving the theoretical models (muffin-tin and tight-binding) are available upon request from the corresponding author.

**Acknowledgements.** We thank Anouar Moustaj for useful insights about the underlying symmetry of the fractal and for helping to setup the model to perform the multi-orbital tight-binding calculations. We are also grateful to Malte Röntgen for insightful discussions about latent symmetry and to Tarik Cysne for useful discussions about spin-orbit coupling in honeycomb lattices. R.C.N, L.E. and C.M.S. acknowledge the research program "Materials for the Quantum Age" (QuMat) for financial support. This program (registration number 024.005.006) is part of the Gravitation program financed by the Dutch Ministry of Education, Culture and Science (OCW). RA thanks the Knut and Alice Wallenberg Foundation for financial support. Authors from SJTU thank the Ministry of Science and Technology of China (Grants No. 2019YFA0308600, 2020YFA0309000), NSFC (Grants No. 11790313, No. 92065201, No. 11874256, No. 11874258, No. 12074247, No. 12104292, No. 12174252 and No. 11861161003), the Strategic Priority Research Program of Chinese Academy of Sciences (Grant No. XDB28000000), the Science and Technology Commission of Shanghai Municipality (Grants No. 2019SHZDZX01, No. 19JC1412701, No. 20QA1405100) and the China Postdoctoral Science Foundation (Grant BX2021184) for financial support.

# Supplementary Material: Topological edge and corner states in Bi fractals on InSb

R. Canyellas[1,2*], Chen Liu[3*], R. Arouca[4], L. Eek[1], Guanyong Wang[3], Yin Yin[3], Dandan Guan[3,5], Yaoyi Li[3,5], Shiyong Wang[3,5], Hao Zheng[3,5], Canhua Liu[3,5], Jinfeng Jia[3,5#], C. Morais Smith[1]

[1]*Institute for Theoretical Physics, Utrecht University, Princetonplein 5, 3584CC Utrecht, The Netherlands*

[2]*Institute for Molecules and Materials, Radboud University, Heyendaalseweg 135, 6525AJ Nijmegen, The Netherlands*

[3]*Key Laboratory of Artificial Structures and Quantum Control (Ministry of Education), Shenyang National Laboratory for Materials Science, School of Physics and Astronomy, Shanghai Jiao Tong University, Shanghai 200240, China*

[4]*Department of Physics and Astronomy, Uppsala University, Uppsala, Sweden*

[5]*Tsung-Dao Lee Institute, Shanghai Jiao Tong University, Shanghai 200240, China*

∗ shared first authors

# corresponding author

## CONTENTS

# I. INTRODUCTION TO FRACTALS

Fractals are self-similar structures, which usually have a non-integer dimension. The Hausorff dimension defines a fractal by how the hypervolume of a structure changes by a scaling transformation. Multiplying a linear scaling of this structure by $b$, the hypervolume of the fractal changes as $b^{d_H}$, where $d_H$ is the Hausdorff fractal dimension. For the Sierpinski triangle, doubling its linear lengths increases its area by a factor of three. Therefore, we have that $d_H = \log_2(3) \approx 1.58$.

A true self-similar structure only occurs for an infinite system. In real systems, one needs to use an approximate definition of a fractal in a finite lattice, which is called a physical fractal. We use the concept of the generation $G$ of a fractal, illustrated in Fig. 1 for the Sierpinski triangle. The process of creating a self-similar structure is made by piercing holes through each triangle in an iterative manner. Notice that the change of the red region in Fig. 1 going from $G = 1$ to $G = 2$ is the same as the change of the whole triangle from $G = 0$ to $G = 1$.

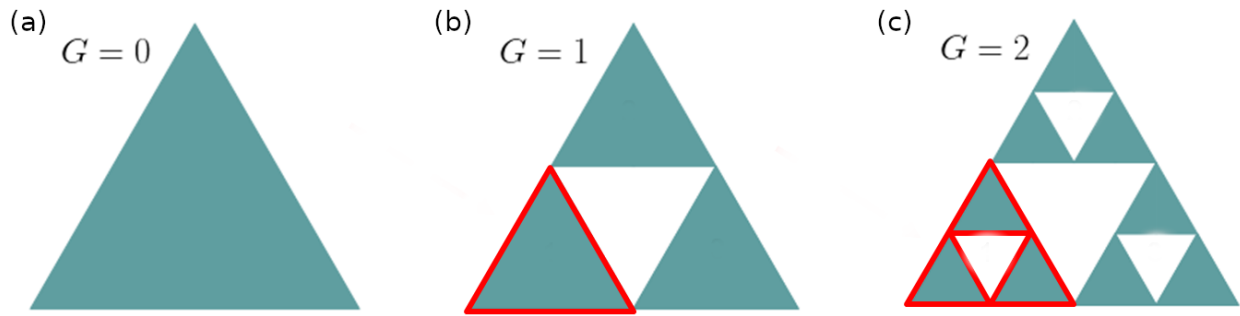

**(a)** $G = 0$  **(b)** $G = 1$  **(c)** $G = 2$

FIG. 1. Construction of a Sierpinski triangle. (a) The zero-th generation, $G = 0$, is a regular triangle with area $A_0$ and perimeter $P_0$. (b) The first generation, $G = 1$, is obtained by removing an inverted triangle from the center of the large triangle. (c) The second generation, $G = 2$, is obtained by removing a triangle in each of the triangles of the first generation, as illustrated by the red line. Higher generations can be obtained by repeating this process in the smaller triangles. Notice that the area $A_2$ of $G = 2$ is now $(3/4)^2$ of $A_0$ and the perimeter $P_2$ of $G = 2$ is $(3/2)^2$ of $P_0$.

As we increase the generation of this fractal, its area diminishes by a factor of $3/4$, while its perimeter increases by a factor of $3/2$. Therefore, in the limit of a true fractal ($G \to \infty$), the area goes to zero and the perimeter goes to infinity. Hence, one can understand why physical fractals appear so constantly in nature: they optimize the area/volume ratio.

Despite being abundant in nature, fractals are not common in solid-state physics because usually van der Waals forces, which are volumetric, dominate the formation of solids. Since fractals optimize the perimeter-to-area ratio, they are favored for materials with stronger covalent bonds, as demonstrated in Ref. [1]. Although one does not usually find such materials in nature, one can build a quantum version of fractals by design. This is precisely what was done in Ref. [2], where the authors built a Sierpinski gasket by placing CO molecules (which act as potential barriers) on a Cu(111) surface (which behaves as a free electron gas above a threshold energy). In this way, they obtained the first quantum fractal. The fractal dimension was not only observed in the lattice but in the wavefunction itself, indicating that the quantum properties of this material are sensitive to the non-integer dimension.

After the original work of Ref. [2], fractals were realized in other kinds of metamaterials, such as photonics [3, 4] and acoustics [5–7]. Ref. [3] described the quantum walk of photons in lattices of waveguides arranged in a fractal shape (Sierpinski gasket, Sierpinski carpet and dual Sierpinski carpet). The authors demonstrated that the exponent of diffusion is not only different than in a system with integer dimension, but also differs from the prediction of transport in a *classical* fractal.

The field progressed further when fractals exhibiting topological states were experimentally realized in metamaterials [4, 6, 7]. In Ref. [4], a periodic potential in a waveguide was used to construct a photonic Floquet topological insulator. Chiral edge states were observed for the same parameter range where there are quantized local Chern numbers. Interestingly, the inner and outer edges circulate in opposite directions (counter and clockwise, respectively). A different route was taken in Refs. [6, 7], where two independent groups realized a 2D Su-Schrieffer-Heeger (SSH) [8] and a variation of the Benalcazar-Bernevig-Hughes (BBH) model [9, 10] in a Sierpisnki carpet. The BBH model is the first example of a model with higher-order topological modes, where the topological states are not present on the entire boundary, but only on parts of it. Specifically for the BBH, there are zero-energy protected states on the corners of the 2D system. Indeed, these acoustic metamaterials host corner states when the quadrupole momentum, a topological invariant, is quantized. For the fractals, one observes topological states in both the inner and outer corners of the system, as well as edge modes at different energies [6, 7]. Moreover, the co-dimension of the outer corner modes was shown to be non-integer [6], and different from the inner ones, thus breaking the general believe that topology must be associated to integer dimensions. Here, we go a step further and investigate fractals in *real materials*.

## II. SAMPLE PREPARATION AND STM MEASUREMENT

We prepared the samples by growing Bi monolayers on an InSb substrate. All experiments were realized using a commercial ultra-high vacuum (UHV) molecular-beam-epitaxy scanning tunneling microscopy (STM) system at a base pressure less than Torr, and all experimental data were measured at the liquid helium temperature of about 4 K. The cleaned InSb(111) substrate surfaces were obtained by about 3 cycles of Ar+ ion sputtering and annealing at about 0.5 h. Fig. 2 (a) shows the topography of the prepared InSb(111)B surface with steps and Fig. 2 (b) depicts the InSb(111)B surface ($3 \times 3$)-reconstructed obtained from Reflection High Energy Electron Diffraction (RHEED). Fig. 2 (c) shows the atomically resolved STM images for the same region under different bias energies which reveal the two types of In-Sb hexamers [11]: $\alpha \wedge \beta$-ring.

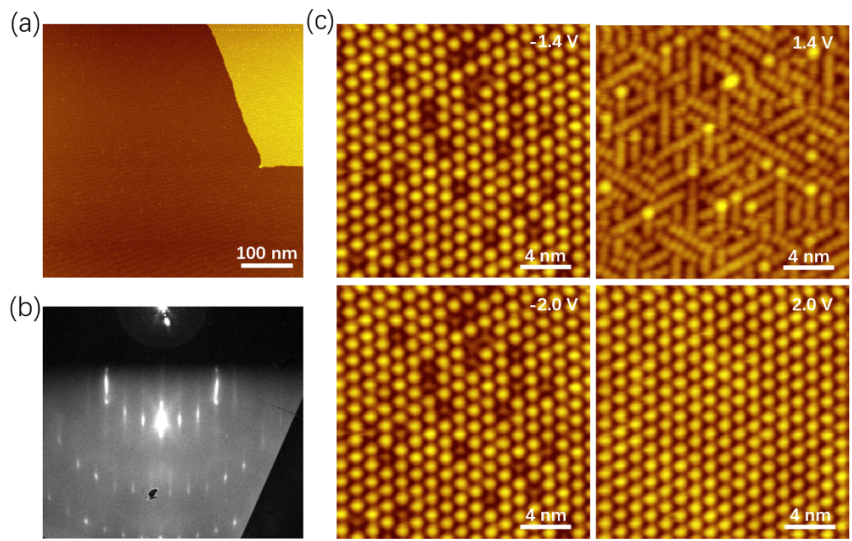

FIG. 2. The prepared InSb(111)B substrate. (a) STM image for topography of ($3 \times 3$)-reconstructed InSb(111)B surface. (b) RHEED pattern of ($3 \times 3$)-reconstructed InSb(111)B surface. (c) Atomically resolved STM images for the same region of ($3 \times 3$)-reconstructed InSb(111)B surface under different bias energy.

Then, we deposit high-purity Bi(99.999%) atoms on the prepared $(3 \times 3)$-reconstructed InSb(111)B substrate surface to obtain the Sierpinski triangle Bi monolayer samples. Here, we use monoatomic layers(MLs) as unit of the coverage, where 1 ML means that all lattice sites in one Bi(111) monatomic layer are occupied by Bi atoms. The deposition time is about 5.5 min, and the deposition rate is about 0.003 ML/s. The range of deposition temperature is around 423 K. After deposition, we quench the system to room temperature with about 1 h cooling time, and finally, we obtain the Sierpinski triangle Bi monolayers on the InSb(111)B surface. STM images in Figs. 3 (a), (b), and (c) show the Sierpinski triangle Bi films formed on InSb(111)B substrate at 1 ML, where different film height levels, 0, $S_1$, $S_2$, and $S_3$ from lower to higher, are labeled [1]. Geometric disorder appears in these STM images, where Sierpinski triangles with different generations connect with each other. By following the red line in Fig 3 (a), one observes the height profiles shown in Fig 3 (d). The distance between different height levels is consistent with previous work [1].

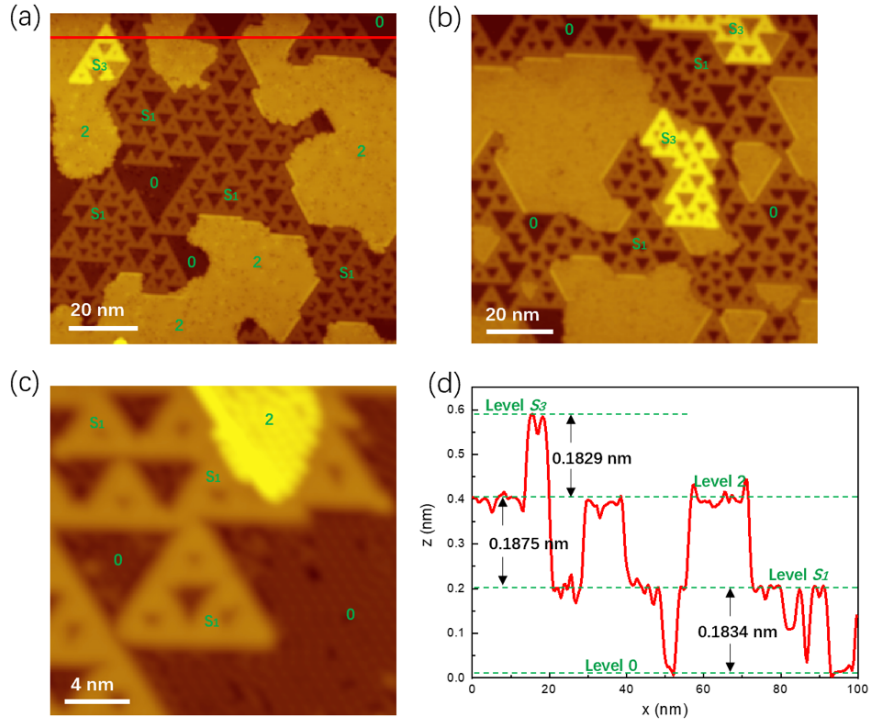

FIG. 3. The topography of Sierpinski triangle Bi monolayers. (a), (b) and (c) STM images of Sierpinski triangle Bi monolayers formed on InSb(111)B substrate at 1 ML. (d) Height profiles for the red line in (a).

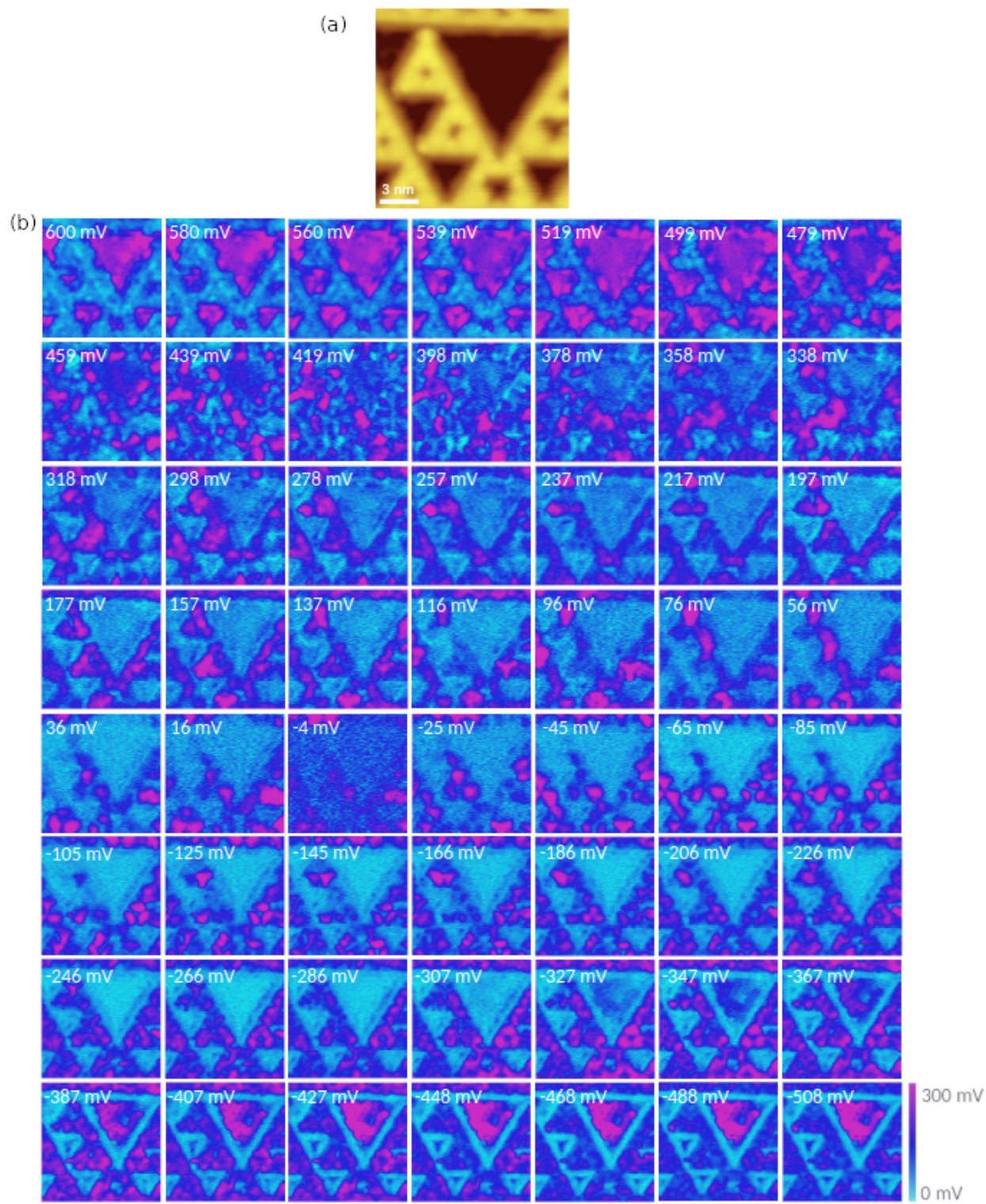

FIG. 4. STS maps. (a) STM image of Sierpinski triangle Bi monolayers. (b) STS maps under different bias energies in the same region as (a). The bias voltages are indicated in each plot.

## III. STS MAPS UNDER DIFFERENT BIAS ENERGIES

In our experiments, we change the bias voltage and obtain the corresponding maps by STM, which show the density of states distribution. Here, we present additional (scanning tunneling spectroscopy) STS maps to support the theoretical results. Fig.4 (a) shows the topography of Sierpinski triangle Bi monolayers formed on InSb(111)B substrate. Fig.4 (b) shows STS maps performed on the Sierpinski triangles at different bias energies. The value of the bias voltage is indicated in each plot.

## IV. THEORETICAL DESCRIPTION

### A. The model

To describe the experimental measurements, we need to simulate the behavior of the electrons when they are confined to move in a Sierpinski gasket. As illustrated in Fig. 5, one can obtain an effective lattice by describing a potential landscape that forces electrons to occupy the Sierpinski triangle regions (green area in Fig. 5 (a)). This method was used in Ref. [2] to engineer an electronic fractal using CO barriers in a 2D electron gas formed at the surface of Cu(111). The purple dots in Fig. 5 (b) represent the electronic lattice sites and the red dots surrounded by a light blue circle in Fig. 5 (c) represent the high potential barriers, which act as scatterers to the electrons, thus creating voids in the 2D electron gas. Here, we investigate a real material, in which the fractal structure is spontaneously formed upon growing Bi on InSb. Nevertheless, the remarkable agreement of the simulations with the experimental data will a posteriori justify that one can use this method to simulate the behaviour of electrons in the spontaneously formed Sierpinski gasket.

The procedure is very similar to a particle in a box, as it consists of solving the one-electron time-independent Schrödinger equation, but with a specific potential landscape that confines the electrons. To reproduce the behaviour of the surface states, we approximate the electrons as free particles, neglecting interactions that they could have among themselves and with the electrons of the bulk. They form a two dimensional free electron gas, which can move in the continuum set of points of the substrate. In addition, we take into account the effects of the intrinsic spin-orbit coupling (ISOC) and the Rashba SOC, and investigate three different cases, depending on the values set to the parameters controlling their relative strength ($\lambda_{\text{ISOC}}$ and $\lambda_{\text{Rsh}}$). A purely kinetic problem ($\lambda_{\text{ISOC}} = 0$ and $\lambda_{\text{Rsh}} = 0$), the case that might be relevant to understand the Bi fractals, when the kinetic and ISOC have a similar order of magnitude ($\lambda_{\text{ISOC}} = 10^6$ and $\lambda_{\text{Rsh}} = 0$), and the case when all of them have a similar order of magnitude ($\lambda_{\text{ISOC}} = 10^6$ and $\lambda_{\text{Rsh}} = 10^9$). This last case helps to probe whether the features obtained in presence of ISOC should be topological or not since the Rashba term is expected to destroy the topological features.

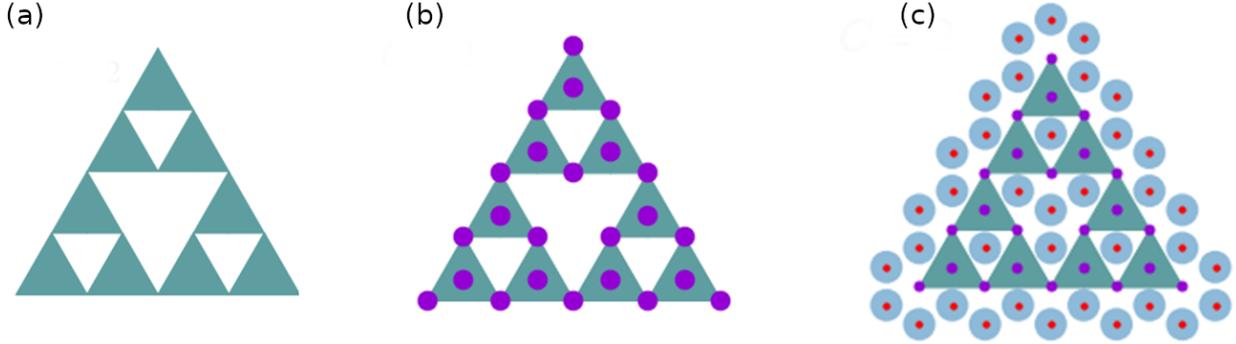

FIG. 5. Realization of a quantum Sierpinski gasket. (a) The structure must be designed such that electrons occupy the Sierpinski gasket (green area). (b) One can build a discrete lattice (purple circles), where the electrons are tightly bound to the atomic sites. (c) Alternatively, one can build a forbidden region of potentials (blue circles centered around the red dots), such that the electrons tend to occupy the Sierpinski area.

The eigenvalue problem that we need to solve is then given by

$$H\boldsymbol{\psi} = \left\{ -\frac{\hbar^2}{2m_e^*}\boldsymbol{\nabla}^2 - \frac{i\lambda_{ISOC}\hbar^2}{(2m_e^*c)^2}\left[(\boldsymbol{\nabla}\mathbf{V})\times\boldsymbol{\nabla}\right]\cdot\boldsymbol{\sigma} - \frac{i\lambda_{\text{Rsh}}\hbar^2}{a_0 m_e^*}(\boldsymbol{\nabla}\times\boldsymbol{\sigma})\cdot\hat{z} + \mathbf{V} \right\}\boldsymbol{\psi} = E\boldsymbol{\psi}. \quad (1)$$

with $V = \text{diag}(v(\boldsymbol{r}))$ the patterned potential for a given generation G (using a muffin-tin or Gaussian potential $v(\boldsymbol{r})$), $m_e^* = m_{eff} \cdot m_{el}$ the product of the electron effective mass and the electron rest mass, $\boldsymbol{\sigma}$ the vector of Pauli matrices, $c$ the speed of light, and $a_0$ the lattice parameter. Going to higher generations does not change much the LDOS due to the self-similarity, but significantly increases the computation time, as the lattice grid needs to be increased. We will therefore restrict our calculations to the lower three generations. Notice that the energies are in units of eV and the lattice is in nm.

To diagonalize the Hamiltonian, we must discretize the space, such that one can approximate the first and second derivative by using the second-order central difference approximation, with periodic boundary conditions. When diagonalizing the Hamiltonian, we set a cutoff to the number of eigenvalues and eigenvectors that we want to obtain. As we are interested in the low-energy spectrum, we solve only for the smallest values. By restricting this set and using sparse matrices, the computation time is highly reduced. After obtaining the set of energies and wavefunctions, we compute the local density of states (LDOS). This gives a measure of how the electronic cloud is distributed, enabling the comparison with images obtained experimentally with the STM. The LDOS is given by

$$LDOS(\boldsymbol{r}, \epsilon) = \sum_{\epsilon'} |\boldsymbol{\psi}_{\epsilon'}(\boldsymbol{r})|^2 \delta(\epsilon - \epsilon') \approx \frac{1}{\pi} \sum_{\epsilon'} |\boldsymbol{\psi}_{\epsilon'}(\boldsymbol{r})|^2 \frac{b}{(\epsilon - \epsilon')^2 + (b)^2},$$ (2)

where we replaced the delta function by a Lorentzian $L(\epsilon - \epsilon')$ with broadening $b = 0.04$ eV. This is necessary because states with energy $\epsilon$ are broadened due to electron-electron and electron-phonon scattering. Thus, the finite lifetime excitations also contribute to the tunneling current measured by the STM. A more in-depth discussion of the broadening will be provided in Sec. IX.

### B. Muffin tin vs Gaussian potential

#### 1. First-generation Muffin-tin results

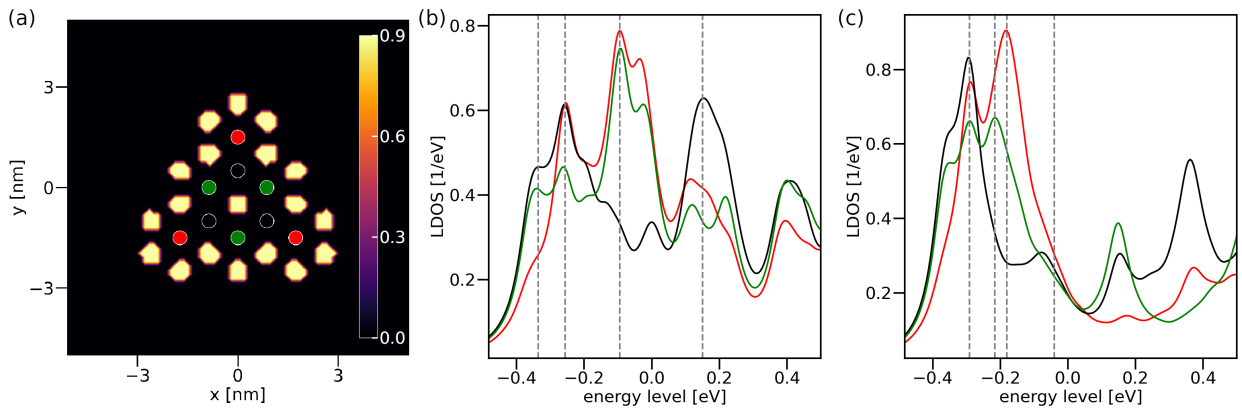

FIG. 6. (a) Simulation box. Scatterers placed at the anti-lattice are represented by yellow shapes with a red contour, which is the zone of non-zero muffin-tin potential. Red, green, and black dots label the sites for the Sierpinski triangle, with connectivity one, two, and three, respectively. (b) Only kinetic and (c) kinetic and ISOC LDOS for all different sites, respectively. Dashed lines represent the energies used in Fig. 7 to obtain the LDOS maps.

First, we will describe the electron system as being confined by a muffin-tin potential. In Fig. 6 (a), the yellow shapes denote the region where the muffin-tin potential is non-zero, while the red, green, and black dots label the sites of the Sierpinski triangle with one, two, and three nearest neighbors, respectively. The asymmetry in the yellow circles is due to the grid resolution. The results shown in Fig. 6 (b) correspond to the mean of the LDOS for the same site color for the case when there is no SOC. The lattice parameter is $a_0 = 1$ nm, the number of grid points has been set for both axes to $n_x = n_y = 80$, and the number of wavefunctions 250. The space between the outer scatterers and the boundary of the square box is set to $2.5 \cdot a_0$, in the abscissa and ordinate axes. If we would not have left enough space, the boundary would have influenced the LDOS inside the Sierpinski. The effective electron mass is $m_{eff} = 0.42$, the onset energy is $u_s = 0.493$ eV, the barrier height is $u = 0.9$ eV, and the effective radius of the scatterers is $r = 0.31$ nm. The onset energy has been chosen in a way to obtain the corner modes at zero energy for the third-generation Sierpinski using the Gaussian potential, as it will be discussed later. This is not an ideal choice for the muffin-tin, but we use it for consistency. In any case, it is just a horizontal shift.

The vertical grey dashed lines in Fig. 6 represent interesting values of energies, at which the LDOS maps shown in Fig. 7 are obtained. Between these lines, the system is evolving from one configuration to another, exhibiting less defined features in the LDOS map. We have chosen the values of energy where the LDOS displays a peak or a dip, and the LDOS map shows a relevant feature. At the first selected energy $E = -336$ meV, the black and green LDOS are displaying a small peak, while the red is still growing, showing that first the bulk is populated. At $E = -256$ meV, all sites have a peak, although the intensity at the green site is smaller. At $E = -95$ meV, one observes a non-bonding configuration, and at $E = 151$ meV the black sites acquire a larger LDOS, while the red and green sites have the LDOS more towards the boundary, with a node between the black and red sites. Those correspond to higher-order states, which will not be discussed here.

We have studied the confinement of a free electron gas by a muffin-tin potential patterning the anti-lattice, and now we move to the main questions: How is the effect of the ISOC in a fractal? Does it open topological gaps and drives matter to topological phases, hosting edge or corner states? To answer these questions, we follow the same procedure as before, but now we will include the ISOC term into the Schrödinger equation by setting $\lambda_{\text{ISOC}} = 10^6$. This value is selected to make both the kinetic and ISOC terms comparable in strength.

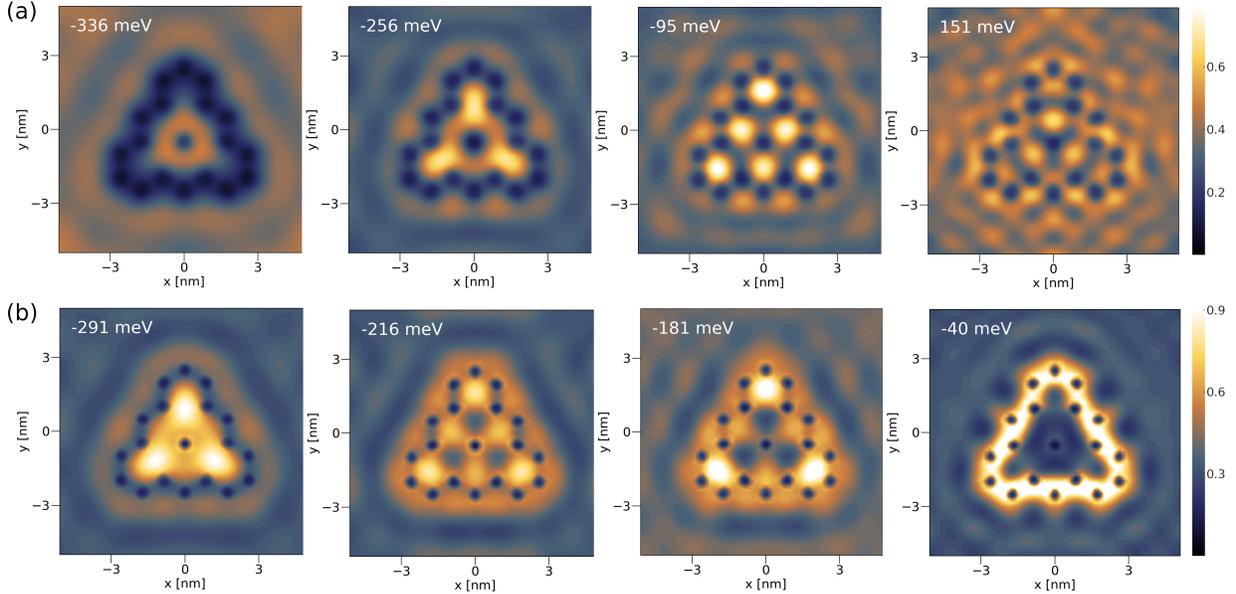

FIG. 7. Maps of the LDOS on the fractal lattice. The figures correspond to the selected values of energy from figure 6 (a) kinetic and (b) kinetic and ISOC, respectively. The maximum of the color bar has been set equal to the LDOS peak of the red site in (a) and (b), respectively.

In Fig. 6 (c), we display the LDOS when the ISOC is present. As before, we have selected a few interesting values of energy (see dashed lines). By looking at the shape of the lines, we can already see that some features have completely changed. First, the original peaks have been shifted to lower energies. The black line now has only one peak at low energy. In addition, the first two configurations at $E = -291$ meV and $E = -216$ meV look similar to the second and third top phases of the kinetic term. However, the ISOC is spreading the LDOS towards the scatterers, to the region where the gradient of the potential is higher. This is an spurious effect, which will be improved later. Nevertheless, at $E = -216$ meV the system does not display a non-bonding configuration as strong as before at $E = -95$ meV, where we had observed two peaks and one valley. The ISOC has decreased the green LDOS while intensifying the red peak, and consequently it is destroying the non-bonding phase and isolating the corner state at $E = -181$ meV.

For the last selected energy, is better to look directly at the LDOS maps shown in Fig. 7 (b), since the LDOS is very low at each site. At $E = -40$ meV, a high (white) LDOS intensity appears surrounding the scatterers of the outer perimeter. For this bias voltage, it seems that the ISOC is producing a high LDOS at the region were the gradient of the potential is higher. In a real material, one can imagine that there is an approximately constant potential at the bulk. When one approaches the edge of the sample, there is a huge change in the potential because it needs to connect with the vacuum. Hence, the boundary is where the gradient of the potential is stronger, and therefore we can think that this phase, which has all the LDOS surrounding the lattice, is an edge state, with no population at the lattice. A topological edge state driven by the ISOC is indeed expected to arise in this system. However, to avoid spurious effects due to the unrealistic abrupt change of the muffin-tin potential at the boundaries, which jumps from zero to $u$ and leads to an infinite derivative, we now investigate the effect of a smoother potential.

## 2. First-generation Gaussian results

In this subsection, we consider again the same system, the first generation of a Sierpiński triangle, but now we replace the shape of the muffin-tin potential by a Gaussian function given by

$$v_G(\boldsymbol{r}) = \sum_i \frac{u}{\Delta\sqrt{2\pi}} e^{\frac{-(\boldsymbol{r}-\boldsymbol{R}_i)^2}{2\Delta^2}}, \tag{3}$$

using the same potential height $u = 0.9$ eV. $\boldsymbol{R}_i$ denotes the position of the $i$-th scatterer. By fixing the full width at half maximum (FWHM) to $d = 0.62$ nm, we computed the corresponding variance $\Delta = d(2\sqrt{2\ln 2})^{-1}$. The smoothness of the Gaussian potential gradient will be reflected in the LDOS, allowing for more resolution in-between the phases, therefore revealing new interesting states.

The Gaussian potential landscape is depicted in Fig. 8 (a), where now the spherical symmetry of the scatterers is approximated with better accuracy. The electronic sites are shown as red, black and green dots, as before. The diameter of the scatterers has been used to set the FWHW, hence defining the variance of the distribution. In Fig. 8 (b) the LDOS shows the characteristic features of the kinetic term that we discussed for the muffin-tin potential, but now the peaks are much clearer. To avoid repetition, we are going to directly discuss Fig. 9 (a), where one can see the LDOS for the four values of energy represented by dashed lines in Fig. 8 (b). Before doing so, it is worth noticing that despite exhibiting the same trend, the peaks are shifted. The first four plots are very similar to the results obtained using the muffin-tin potential, but now each phase is much sharper because the peaks are more pronounced and smoother. As before, we set the maximum of the color bar

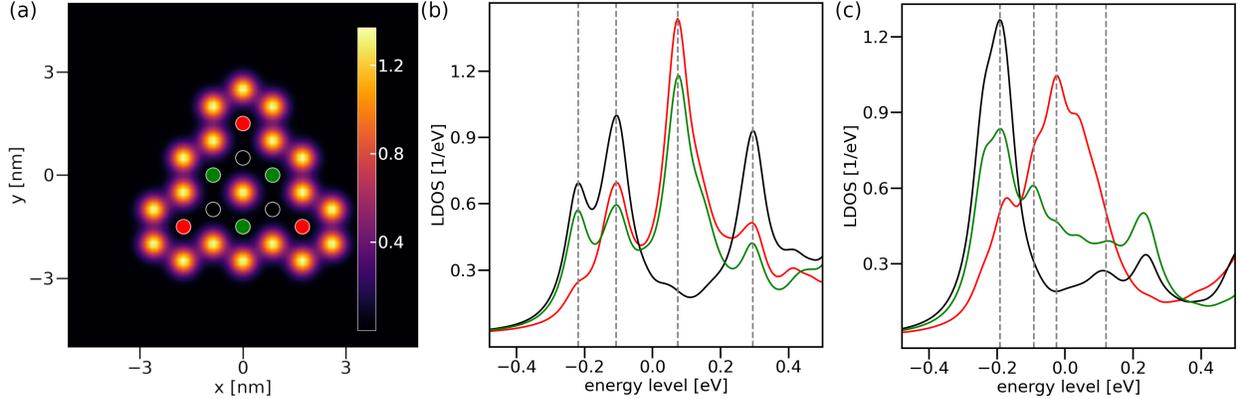

FIG. 8. (a) Simulation box. The FWHM of the Gaussian potential is given by the purple disk diameter encircling a yellow dot. Red, green, and black label the sites for the Sierpinski triangle with connectivity one, two, and three, respectively. (b) Kinetic and (c) kinetic plus ISOC LDOS for all different sites, respectively. Dashed lines correspond to selected the energies at which the LDOS maps will be shown in Fig. 9.

to the maximum of the red peak, to better visualize all the features.

The Gaussian results are not very different from the muffin-tin when we consider only the kinetic term. However, the differences become much more relevant when we include the ISOC, see Figs. 8 (c) and 9 (b). The selected phases to depict the Gaussian potential LDOS maps are similar to the ones that we showed before in Fig. 7 (b).

In Fig. 8 (c), we observe again the main effects of the ISOC term: the breaking of the anti-bonding phase and the appearance of the corner state, which is now much clearer because the difference of the LDOS between the red and the other sites has increased, and the formation of a bright edge state. The scan maps of the LDOS at the selected energy values are shown in Fig. 9 (b). At $E = -192$ meV, there is a peak at the black and green sites, and a smaller peak at the red site. As expected, there are three small self-similar triangles, with a bright spot at the black site. This is a bulk phase. At $E = -92$ meV, the LDOS spreads towards the edge of the lattice. It looks like an edge state (now centered at the boundary sites) that has a bit more occupation at the corners. Upon increasing energy, this trend continues, the population at the green site keeps decreasing, till the corner state is isolated at $E = -26$ meV. Then, at $E = 120$ meV we find an edge state as before, along the outer perimeter.

Figure 10 shows, for each type of site (a) red, (b) green, and (c) black, the trend of the LDOS for different values of the intrinsic ISOC parameter $\lambda_{\text{ISOC}}$ listed at the legend. All sites exhibit a similar behaviour, but the position and intensity of the peaks change slightly. Generally, we see how the ISOC is shifting all peaks to lower energies. Fig. 10 (a) shows how the red peak is isolated around $E = 0$ eV. For $\lambda_{ISOC} = 0$, the green site in Fig. 10 (b) had a at a similar energy as the red site in Fig 10 (a), but upon increasing $\lambda_{ISOC}$ this peak disappears and another one emerges at lower energies. Figure 10 (c) shows how the initial two peaks for the black sites change and become only one at lower energies.

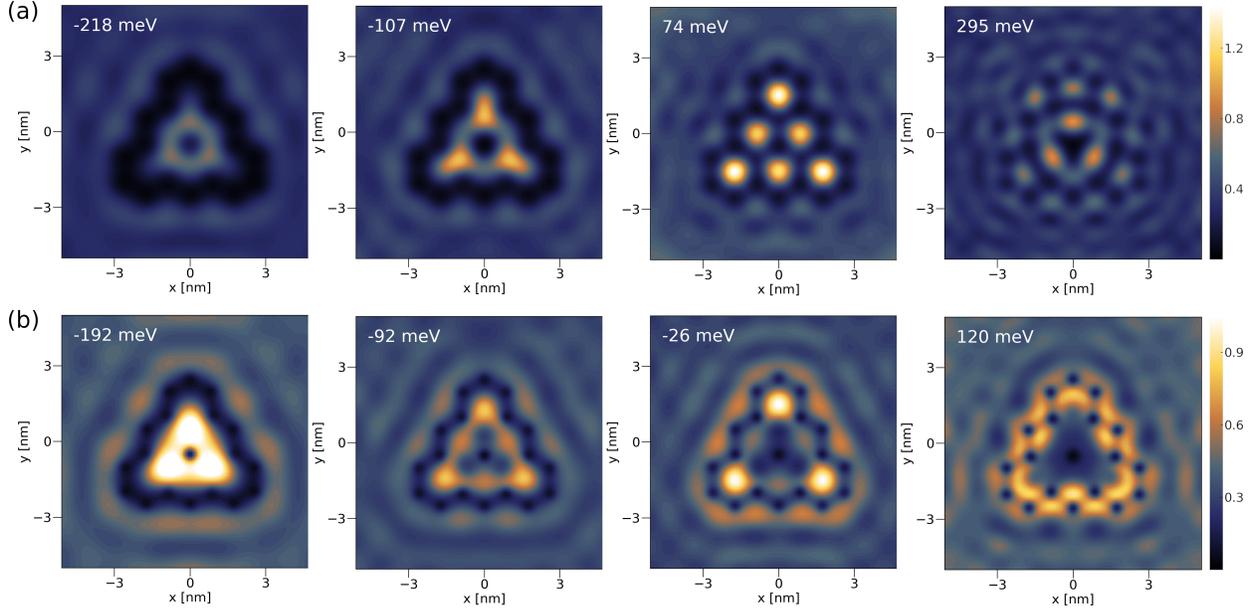

FIG. 9. Maps of the LDOS for the 1st generation Sierpiński structure for the kinetic model (a) and kinetic plus ISOC (b). The maps correspond to the selected values of energy shown in Fig. 8. For each model the maximum of the color bar has been set equal to the maximum of the red peak.

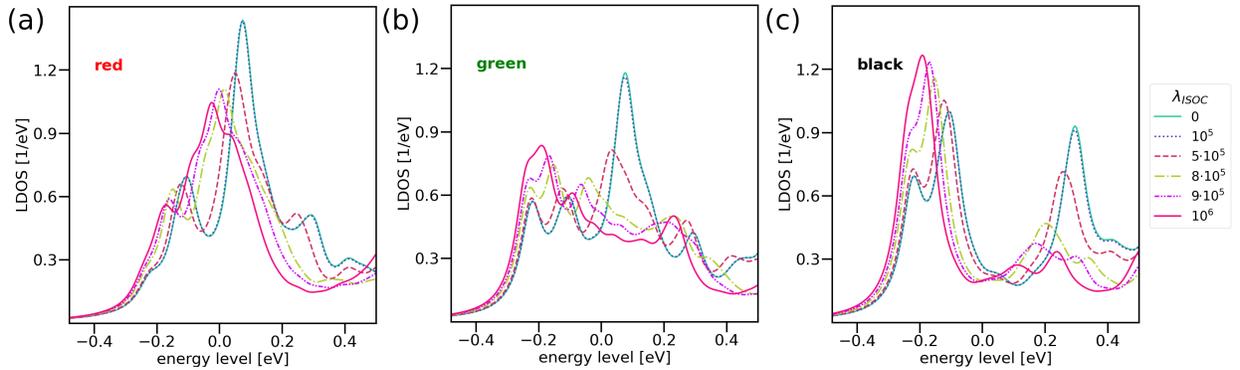

FIG. 10. LDOS at each site for simulations with different values of $\lambda_{ISOC}$. The site color is indicated in the left upper corner of each image. It follows the color code of Fig. 6, with (a) representing the corners (red), (b) the edge (green), and (c) the bulk (black).

## V.  RESULTS FOR THE SECOND GENERATION

The Sierpinski triangles obtained experimentally seem to be of second generation. However, in the main text we showed theoretical results for third generation since: (i) the phases observed should be qualitatively similar due to the self-similarity of the Sierpinski triangle, and (ii) the third generation fractal has less finite-size effects. In this Section, we show the phases obtained for the second generation Sierpinski triangles and compare the results with the third generation ones shown in Fig.2 of the main text. The results are displayed in Fig. 11. The selected energy values of the maps vary slightly with respect to the ones chosen in Fig. 2 of the main text because we choose them according to the peaks and dips appearing in Fig. 11 (a), (f), and (k). Similarly to the main text, the columns represent the values of the parameters used in the simulation. The first (pink) column [Figs. 11 (a)-(e)] shows results for only the kinetic term ($\lambda_{\text{ISOC}} = \lambda_{\text{Rsh}} = 0$). The LDOS exhibits distinct peaks for the bulk modes around $E = -216$ meV and $E = -135$ meV. The corresponding maps [Figs. 11 (b) and (c)] indeed reveal bulk states that differ by their connectivity: the one with lower energy has stronger weight inside the first generation Sierpinski triangles, whereas the other one is stronger at the links between these triangles. At around $E = 20$ meV, there is a depletion of the LDOS for all sites [Fig. 11 (d)]. There is a large dip for the bulk states around $E = 95$ meV, which corresponds to the anti-bonding bulk state [Figs. 11 (e)]. Notice that these phases are precisely the ones shown for the 3$^{\text{rd}}$ generation in Figs. 2 a-e of the main text. The results for a finite value of ISOC ($\lambda_{\text{ISOC}} = 10^6$, $\lambda_{\text{Rsh}} = 0$) are shown in the second (purple) column [Figs. 11 (f)-(j)].

We observe a significant change of the LDOS peaks [Fig. 11 (f)] in comparison with the kinetic case. The bulk peak observed around $E = -201$ meV corresponds now to a state occupied uniformly [Fig. 11 (g)]. More importantly, now we observe a clear peak corresponding to corner states close to zero energy [Fig. 11 (h)]. For higher energies (in a region close to $E = 119$ meV), one finds the edge states shown in Figs. 11 (i) and (j).

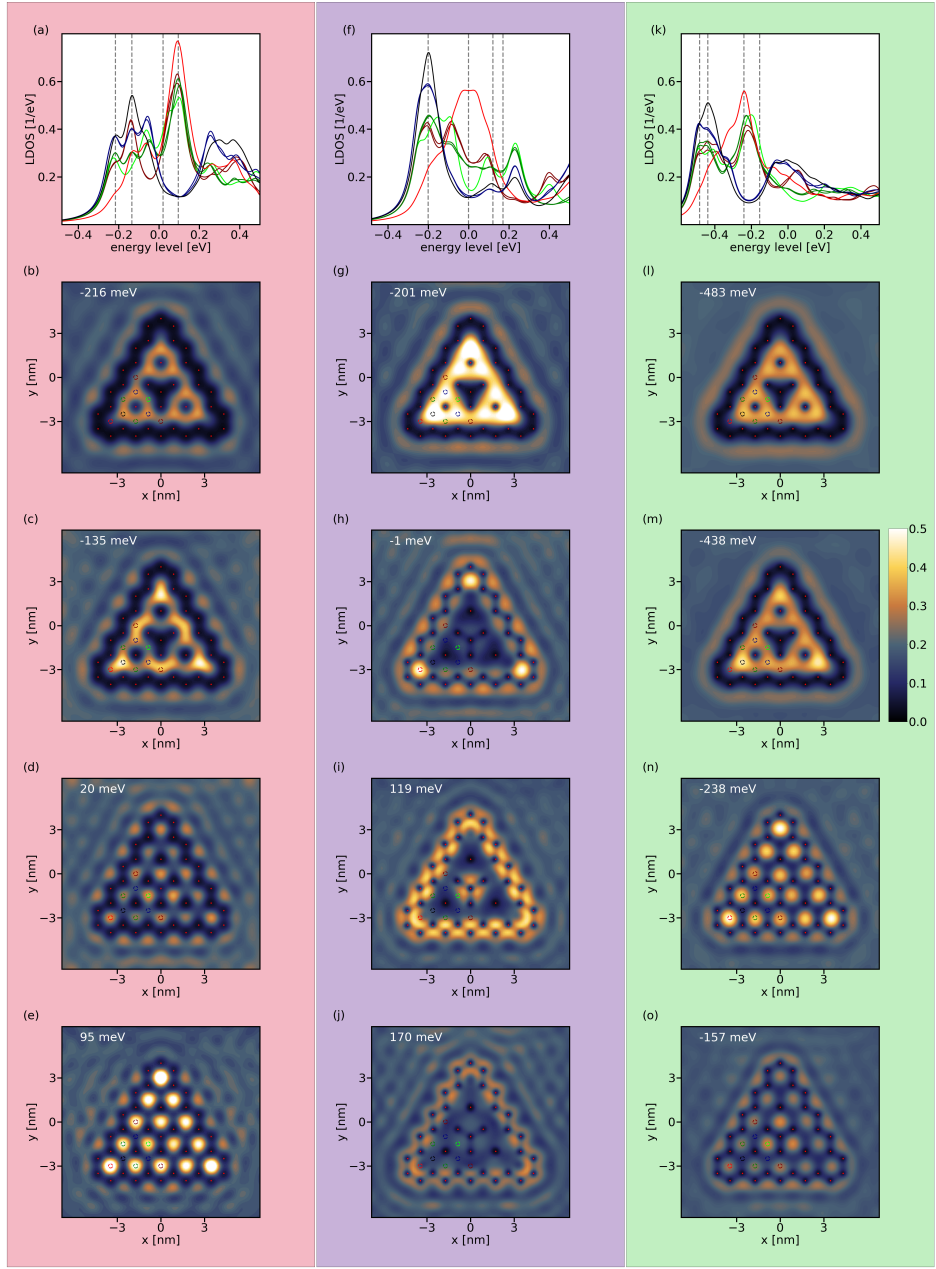

FIG. 11. Theoretical LDOS maps for the second generation (G=2) Sierpinski triangle obtained in the framework of the muffin-tin method with a smooth Gaussian potential. The calculations shown in the first column (pink) include only the kinetic term, while the second column (purple) also includes the ISOC, and the third (green) includes all three terms, kinetic, intrinsic, and Rashba SOC. The number of grid points in both axes is $n_x = n_y = 100$, the number of waves 750, and the Gaussian potential is characterized by the FWHM $d = 0.62$ nm and a potential height $u = 0.9$eV. The electron effective mass $m_{\text{eff}} = 0.42$, the lattice parameter $a_0 = 1$nm, and the intrinsic and Rashba SOC are $\lambda_{\text{ISOC}} = 10^6$ and $\lambda_{\text{Rsh}} = 10^9$, respectively.

Some important changes arise with respect to the 3$^{rd}$ generation result shown in the main text. Although the phases, their orders, and energy scales are approximately the same, the inner edge state that was visible at $E = 185$ meV in Fig. 2 j of the main text does not appear here. Therefore, calculations for the 2$^{nd}$ generation do not allow us to distinguish between the state with high LDOS at the outer edges and the other state with only inner edge states (Figs. 2 i and 2 j of the main text respectively). Upon introducing also a Rashba SOC ($\lambda_{ISOC} = 10^6$, $\lambda_{Rsh} = 10^9$), we obtain phases similar to the purely kinetic case, see Figs. 11 (k-o). The features characteristic of the ISOC are destroyed, thus suggesting that these had a topological origin. Here, we obtain the same phases of Figs. 2 k-o in the main text; hence, there are no differences between the 2$^{nd}$ and 3$^{rd}$ generation results.

## VI. 2D MAPS FOR 3$^{RD}$ GENERATION: RASHBA VS INTRINSIC SOC

In Fig. 2 of the main text (green column), we have shown the LDOS maps at values for which the LDOS curves had a maximum or a minimum after including the Rashba term. Here, we show in Fig. 12(a) version of Fig. 2 where the LDOS maps for the Rashba (green) are taken at the same energy values as for the case with only ISOC (purple). The first column [Figs. 12(a-e)] corresponds to the simulation with both $\lambda_{ISOC} = \lambda_{Rsh} = 0$; the second column [Figs. 12(f-j)] to $\lambda_{ISOC} = 10^6$ and $\lambda_{Rsh} = 0$; and the third column [Figs. 12(k-o)] to $\lambda_{ISOC} = 10^6$ and $\lambda_{Rsh} = 10^9$. Each row corresponds to a different energy. The energy $E \approx -210$ meV [Figs. 12(b, g, l)] is selected because the LDOS with only ISOC [Fig. 12(f)] presents a peak corresponding to the occupation of bulk states. This is verified directly in Fig. 12(g) and is also seen without SOC [Fig. 12(b)]. In the presence of Rashba [Fig. 12(l)], this becomes the anti-bonding state. Then, we analyze $E \approx 0$ meV [Figs. 12(c, h, m)], which corresponds to the corner states in the presence of ISOC [Figs. 12(f, h)]. The corner states are absent without ISOC [Fig. 12(c)] and in the presence of Rashba [Fig. 12(m)]. Similarly, for $E \approx 130$ meV [Figs. 12(d, i, n)], one observes edge states only for the case with ISOC [Fig. 12(i)]. The final value of energy, $E \approx 185$ meV [Figs. 12(e, j, o)], corresponds to edge states in the inner boundary of the Sierpinski triangle for the case of only ISOC [Fig. 12(j)]. These states are not visible for the other parameter values [Figs. 12(e, o)]. We certify then that the conclusion of the main text, i.e. that the corner and edge states are created by the ISOC and destroyed by the Rashba, is confirmed upon selecting fixed energy values.

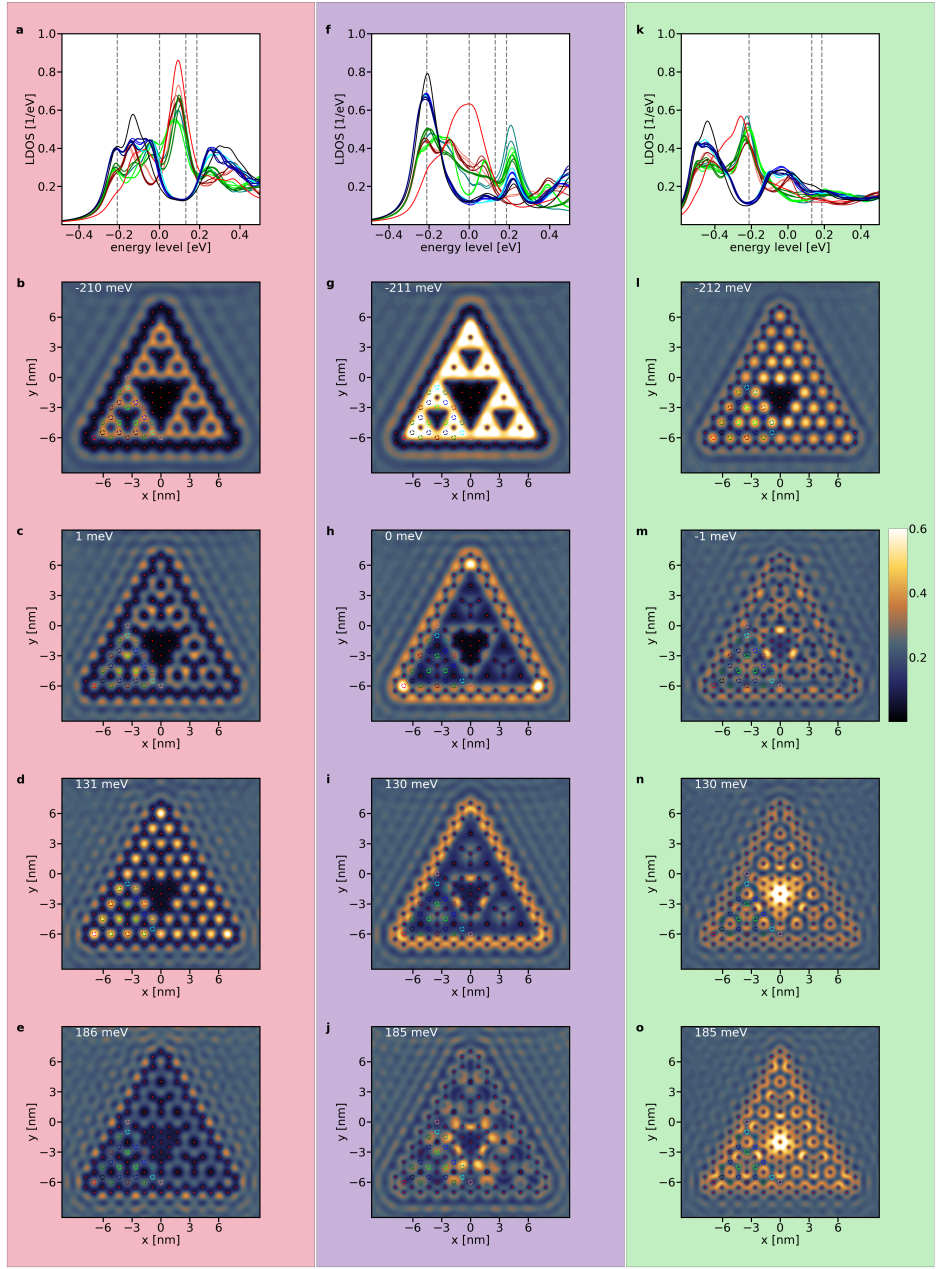

FIG. 12. Theoretical LDOS maps for the third generation (G=3) Sierpinski triangle obtained in the framework of the muffin-tin method with a smooth Gaussian potential. The calculations shown in the first column (pink) include only the kinetic term, while the second column (purple) also include the ISOC, and the third (green) include all three terms, kinetic, intrinsic, and Rashba SOC. The number of grid points in both axes is $n_x = n_y = 200$, the number of waves 1500, and the Gaussian potential is characterized by the FWHM $d = 0.62$ nm and a potential height $u = 0.9$eV. The electron effective mass $m_{\mathrm{eff}} = 0.42$, the lattice parameter $a_0 = 1$nm, and the intrinsic and Rashba SOC are $\lambda_{\mathrm{ISOC}} = 10^6$ and $\lambda_{\mathrm{Rsh}} = 10^9$, respectively.

## VII.   EFFECT OF DISORDER STRENGTH

In this Section, we elaborate on the effect of different disorder strengths for potential and displacement disorders. We show that for a moderate disorder, the conclusion of the main text is kept: i.e., the corner and edge states are unchanged, thus confirming the topological origin. Nevertheless, for strong enough disorder the bulk and edges will eventually be modified.

### A.   Potential disorder

In Fig. 13, we summarize the effect of potential-height disorder on the bulk, corner, and edge states for values of disorder higher than the one presented in Fig. 4 of the main text. The first row of Fig. 13 display the results for no disorder and the second row for 20% $[u = 0.9 \times (1 \pm 0.2)$eV] of disorder. We verify that the bulk states, Figs. 13 (a-e), become slightly asymmetric in presence of disorder, presenting a larger occupancy in the upper and lower left parts of the lattice, while being depleted in the lower right part. The phase with states at corner of the small and larger triangles and also intensity at the middle side of the inner triangles remains discernible despite the disorder, see Fig. 13 (b) and (f). In opposition, the corner states, Fig. 13 (c) and (g), are unchanged and better defined, since the competing edge states that appeared at similar energies are now more depleted. The robustness of the edge states to disorder is also verified in Fig. 13 (d) and (h), where these states are shown to remain intact. In summary, for 20% of disorder, we verify that the bulk states become slightly asymmetric and the corner and edge states remain unchanged.

Above a certain threshold of disorder strenght ($\sim 25\%$, not shown), the results change. We choose the representative values of 30% $[u = 0.9(1 \pm 0.3)]$ eV, third row of Fig. 13) and 50% $[u = 0.9(1 \pm 0.5)]$ eV, fourth row of Fig. 13) of disorder to exemplify this point.

The bulk states are clearly asymmetric and start to become depleted [Figs. 13 (i), (m)]. The states at the center of the inner triangle sides are still visible, but irregular, Fig. 13 (j) and (n). The corner states remain unchanged [Figs. 13 (k) and (o)], while the edge states become more irregular [Figs. 13 (l) and (p)]. With increasing disorder strength, the effects become even more pronounced. It is remarkable that the topological features remain visible despite the enormous value of disorder.

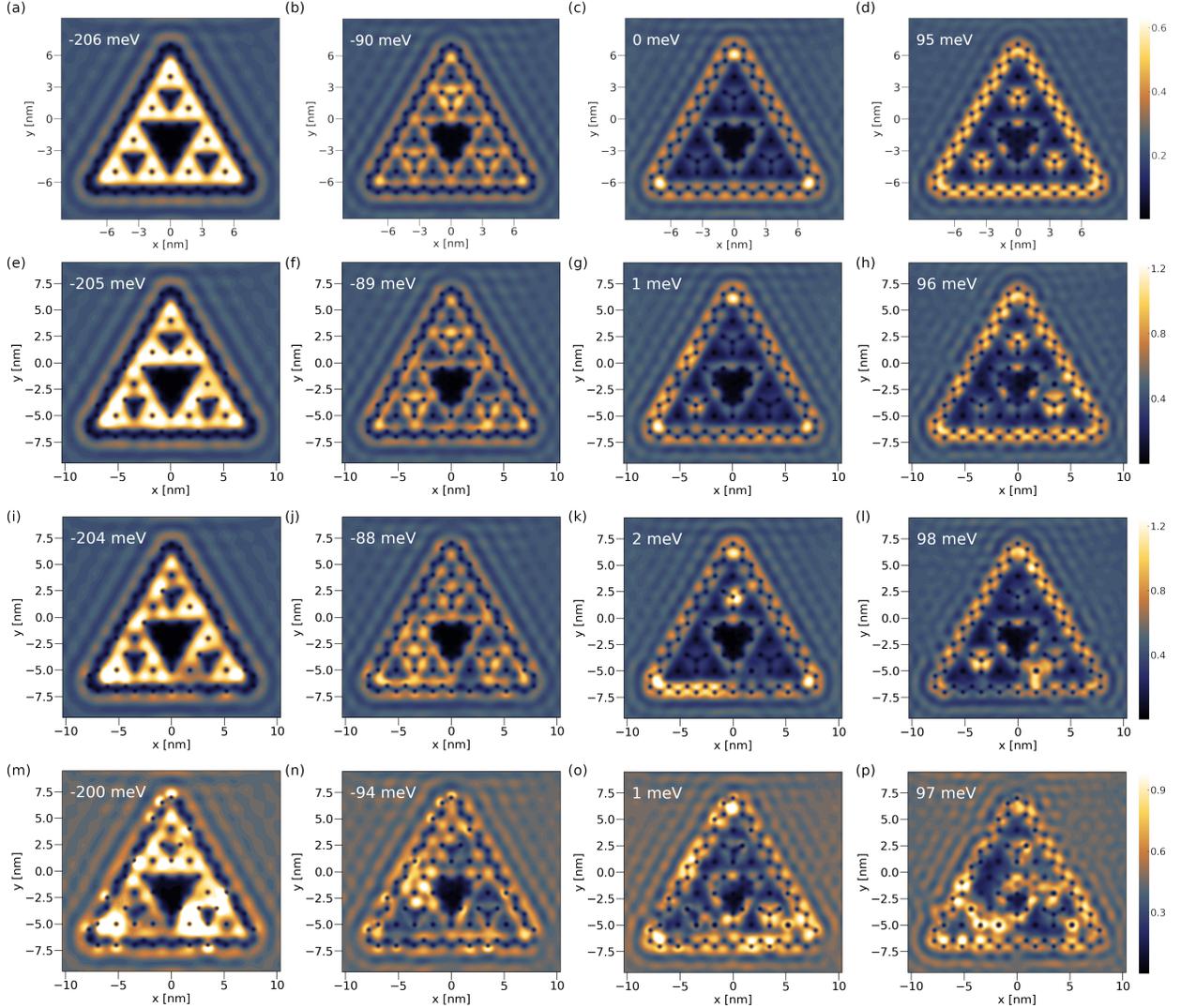

FIG. 13. Theoretical LDOS maps for the third generation (G=3) Sierpinski triangle obtained theoretically in the framework of the muffin-tin method with a smooth Gaussian potential. The height of the Gaussian for each scatterer is taken from a uniform distribution of: (a-d) no disorder,(e-h) $u = 0.9(1 \pm 0.2)$eV, (i-l) $u = 0.9(1 \pm 0.3)$eV, and (m-p) $u = 0.9(1 \pm 0.5)$eV. For each interval, we select representative values of energies indicated in the upper left corner. The number of grid points in both axes is $n_x = n_y = 200$, the number of waves 750, and the Gaussian potential is characterized by the FWHM $d = 0.62$ nm. The electron effective mass $m_{\text{eff}} = 0.42$, the lattice parameter $a_0 = 1$nm, and the intrinsic and Rashba SOC are $\lambda_{\text{ISOC}} = 10^6$ and $\lambda_{\text{Rsh}} = 0$, respectively.

## B. Displacement disorder

The system is more sensitive to displacement disorder. For disorder's strength below about 10%, there is not much change in the patterns. However, for stronger disorder the fractal lattice itself starts to be affected, becoming more amorphous, and the LDOS changes sensibly.

We summarize in Fig. 14 the results for disorder strengths larger than 10%. In Fig. 14 (a)-(d), we plot again the results for no disorder to facilitate the comparison. For 20% [Figs. 14 (e-h)], the bulk states [Figs. 14 (e)] are asymmetric; further, there is an evident depletion of states in the bonds between the fist generation triangles [Fig. 14 (f)]. The corner states are still present [Fig. 14 (g)], but the edge states are destroyed by this disorder [Fig. 14 (h)]. For higher values of disorder, even the holes that characterize the first-generation Sierpinski triangle are now ill-defined, as can be seen in Figs. 14 (m)-(p). This large change in the potential landscapes drastically affects the states observed. For 30% of disorder [Figs. 14 (i)-(l)], the only remaining states are the low energy bulk states [Figs. 14 (i)] and a corner state [Figs. 14 (k)] that is strongly asymmetric, being visible only in the left corner. For disorder strength of 50%, the lattice is destroyed, and we do not observe any clear pattern in the LDOS [Figs. 14 (m)-(p)].

We then conclude that high values of displacement disorder destroy the lattice and, in the limit of an amorphous material, we cannot observe the states discussed in the main text since the fractal lattice itself is not preserved.

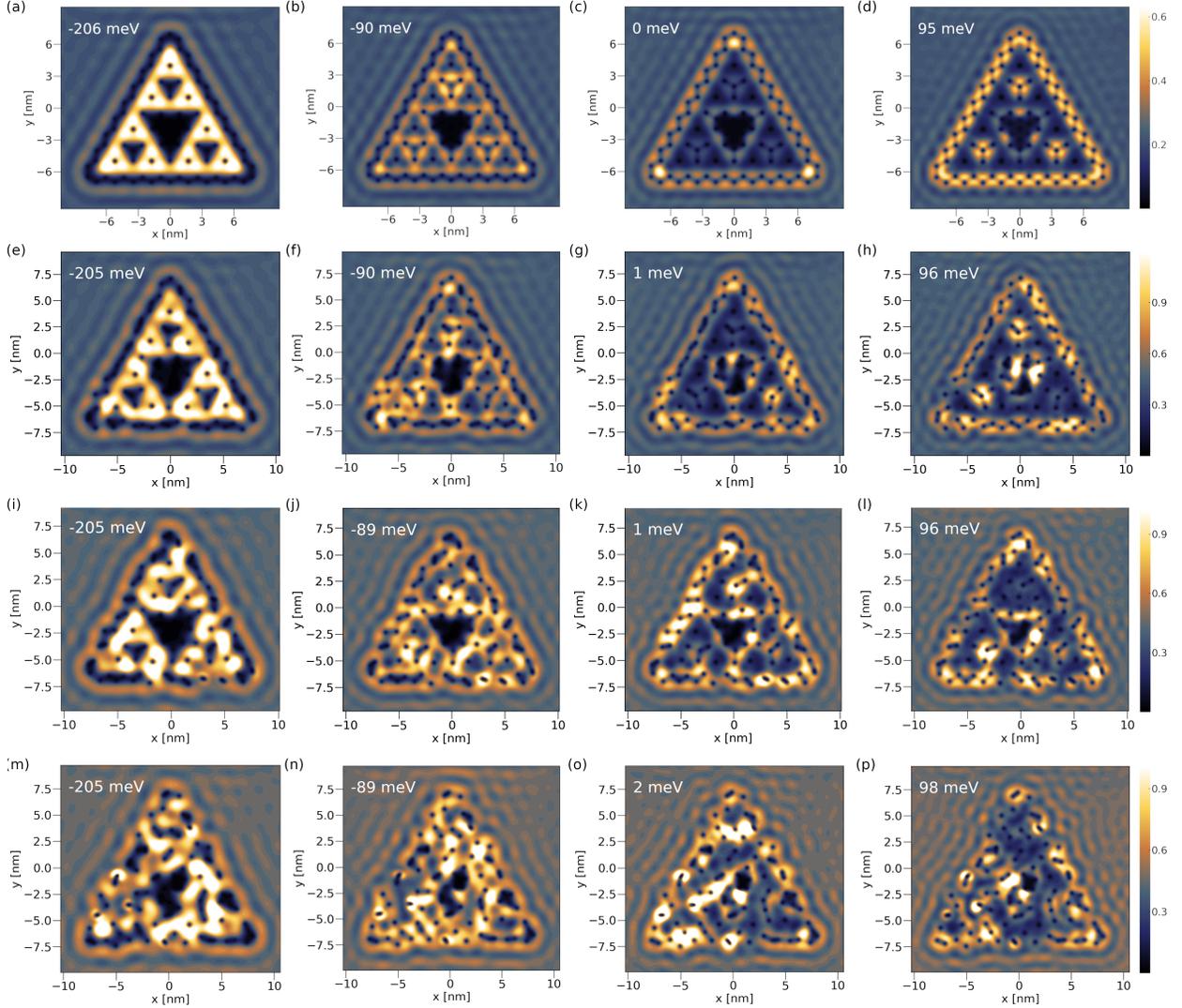

FIG. 14. Theoretical LDOS maps for the third generation (G=3) Sierpinski triangle obtained theoretically in the framework of the muffin-tin method with a smooth Gaussian potential. The distance between the scatterers is taken from a uniform distribution of: (a-d) no disorder, (e-h) $a_0 = (1 \pm 0.2)$ nm, (i-l), $a_0 = (1 \pm 0.3)$ nm, and (m-p), $a_0 = (1 \pm 0.5)$ nm. For each interval, we select representative values of energies indicated in the upper left corner. The number of grid points in both axes is $n_x = n_y = 200$, the number of waves 750, and the Gaussian potential is characterized by the FWHM $d = 0.62$ nm and a height of $u = 0.9$ eV. The electron effective mass $m_{\text{eff}} = 0.42$ and the intrinsic and Rashba SOC are $\lambda_{\text{ISOC}} = 10^6$ and $\lambda_{\text{Rsh}} = 0$, respectively.

## VIII.   DETAILED COMPARISON BETWEEN THEORY AND EXPERIMENTS

### A.   Experimental LDOS maps

Now, we describe the general trends and the different phases distinguished in the full range of energies investigated experimentally. Beginning at the bottom of Fig. 15, for the lowest value of the bias voltage ($E = -507$ meV), we see a sharp cyan line surrounding the external boundary and filling the inner hole. These sharp cyan lines delimiting the contour of the fractal are also visible in Fig. 2 Hence, there is a very low LDOS in these regions. Inside the Sierpinski triangle, there are some triangular pink shapes denoting the high LDOS at the bulk observed in the first-generation Sierpinski triangle. As one increases the voltage, the LDOS becomes more defined, and for $E = -447$ meV, one can even identify the hole of the second generation by a darker region in the center of the three smaller triangles. This corresponds to a bulk phase, where the LDOS is located inside the Sierpinski triangle. The well defined cyan contour around the larger triangle remains from $E = -507$ meV ti $E = -346$ meV. For the next images, there is much disorder, and it is not ve'ry clear what can be identified. Nevertheless, around $E = -266$ meV, a "blobby" phase begins to form. This becomes sharper while increasing the energy, see e.g. the LDOS at $E = -104$ meV, where the top and the left $G = 1$ Sierpinski triangles resemble a three-leaf clover (also for the right bottom one, but not as clear). The three-leaf clover could be related to a corner state of the first generation [12]. Thus, the system could be changing from the bulk phase to a corner state of the previous generation. Increasing the bias voltage further, around $E = 16.4$ meV we observe that the corner states that connect the first generation of Sierpinski triangles have disappeared, and now there are only corner states for the top and left corners of the $G = 2$ Sierpinski triangle (we suppose that the last right-bottom corner is missing due to the disorder). From the corner state, the LDOS begins to spread along the edge, as one can see around $E = 96.9$ meV. This is a topological edge state. The next values of voltage that seem to exhibit some clear features are around E = 218 meV. Now, the LDOS is localized around the inner hole of the Sierpinski triangle. The next images have too much disorder, and the features are not very clear. At most, it seems that the LDOS in pink is differentiating into small self-similar parts surrounding the Sierpinski triangle holes (see image at $E = 318$ meV). Finally, at the highest values of bias voltage (from $E = 439$ meV) onwards, we find again that all the LDOS is outside the Sierpinski triangle, in agreement with our findings.

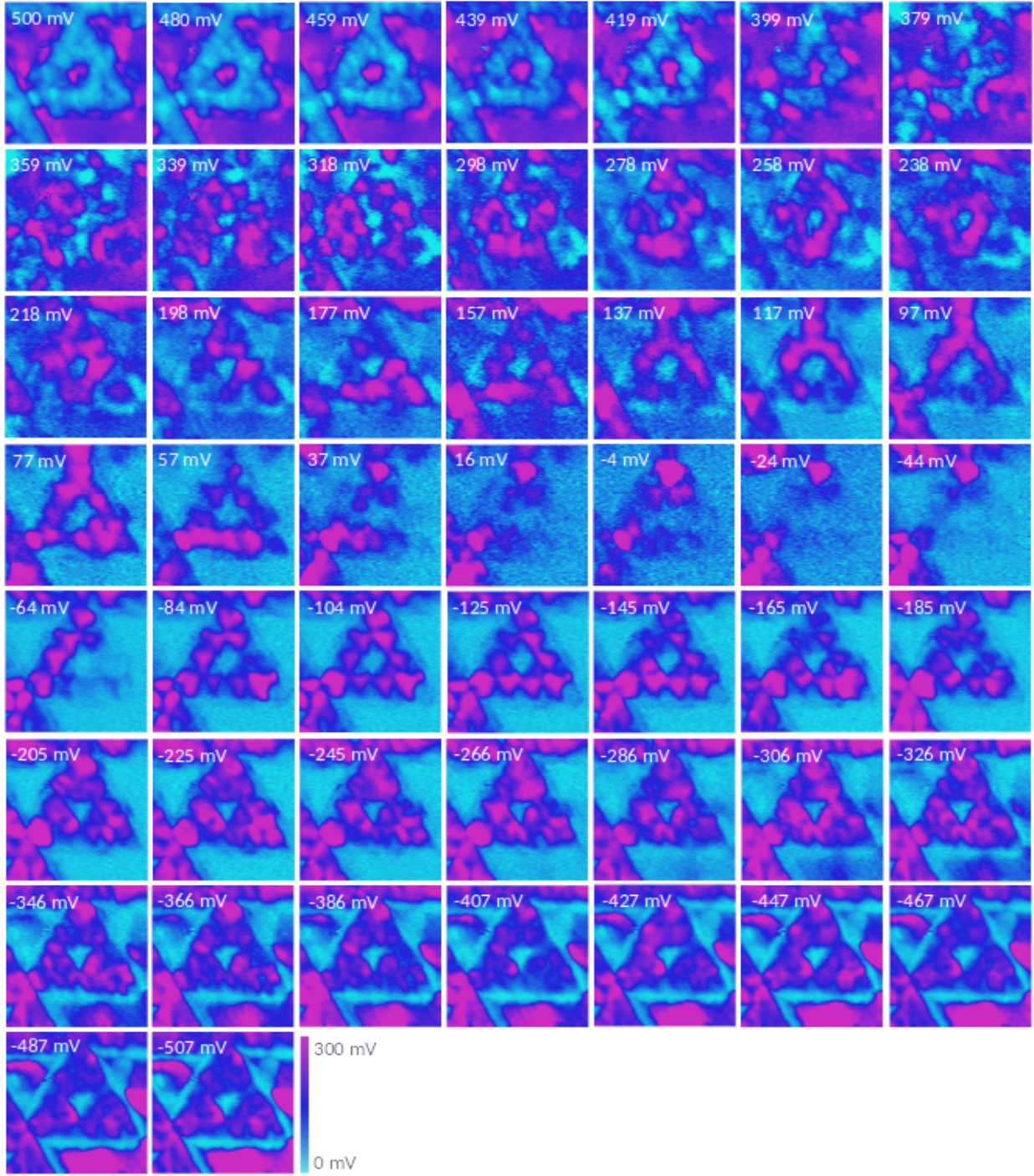

FIG. 15. Full range of experimental LDOS data. Voltage values are indicated in each plot.

## B.  Theoretical muffin-tin LDOS maps

In the main text, we show in Fig. 3 a comparison between the characteristic phases observed in the experimental LDOS maps of Fig 15 and our theoretical results for the muffin-tin model. Here, we depict in Fig. 16 the same results, but now for a much larger set of energies. The parameters used for the simulation are the same as in Fig. 3. At the lowest value of energy (see bottom-right image), $E = -201$ meV, one observess a high LDOS at the bulk while a sharp cyan line surrounds the exterior perimeter and fills the inner holes. A self-differentiation can be seen, as the points connecting the lower generation triangles have a slightly lower intensity. Upon increasing the energy, the pink region becomes less intense and moves towards the edges of the inner triangles as at $E = -140$ meV. At $E = -100$ meV, the bulk starts to be depleted. Around zero energy, the LDOS is decreased over all the fractal, except at the outer corners, which display a high intensity. Afterwards, the LDOS becomes high at the outer edge and at the center of the lines forming the small hole of the first and second generation, as at $E = 70$ meV.

At $E = 120$ meV, the LDOS at the middle point of the first generation triangle lines has faded, but the ones at the second generation remain. The next images exhibit the evolution from the outer to the inner edge that we showed in the comparison of Fig. 3 (see inner-hole edge state at $E = 180$ meV). At $E = 220$ meV, the LDOS is localized surrounding the inner scatterers of the first and second generation. The shapes ressemble "depleated" corner states, in which the corners of the triangle are missing but the adjacent sites have a high intensity. This phase could be connected to the one discussed theoretically in Ref. [6], where trimer states are formed, with zero intensity at the corner. The others values of energy show how the LDOS becomes intense around all the interior scatterers. The main difference is that the LDOS shifts a bit between different regions. For example, at $E = 321$ meV, the LDOS is more pronounced at the inner part of the second generation, while at $E = 466$ meV, it is more intense near the outer boundary. At $E = 501$ meV, the high intensity appears around the second generation scatterers. These features correspond to higher orbitals and will not be discussed here. The last values of energy show how the LDOS at the bulk becomes depleted.

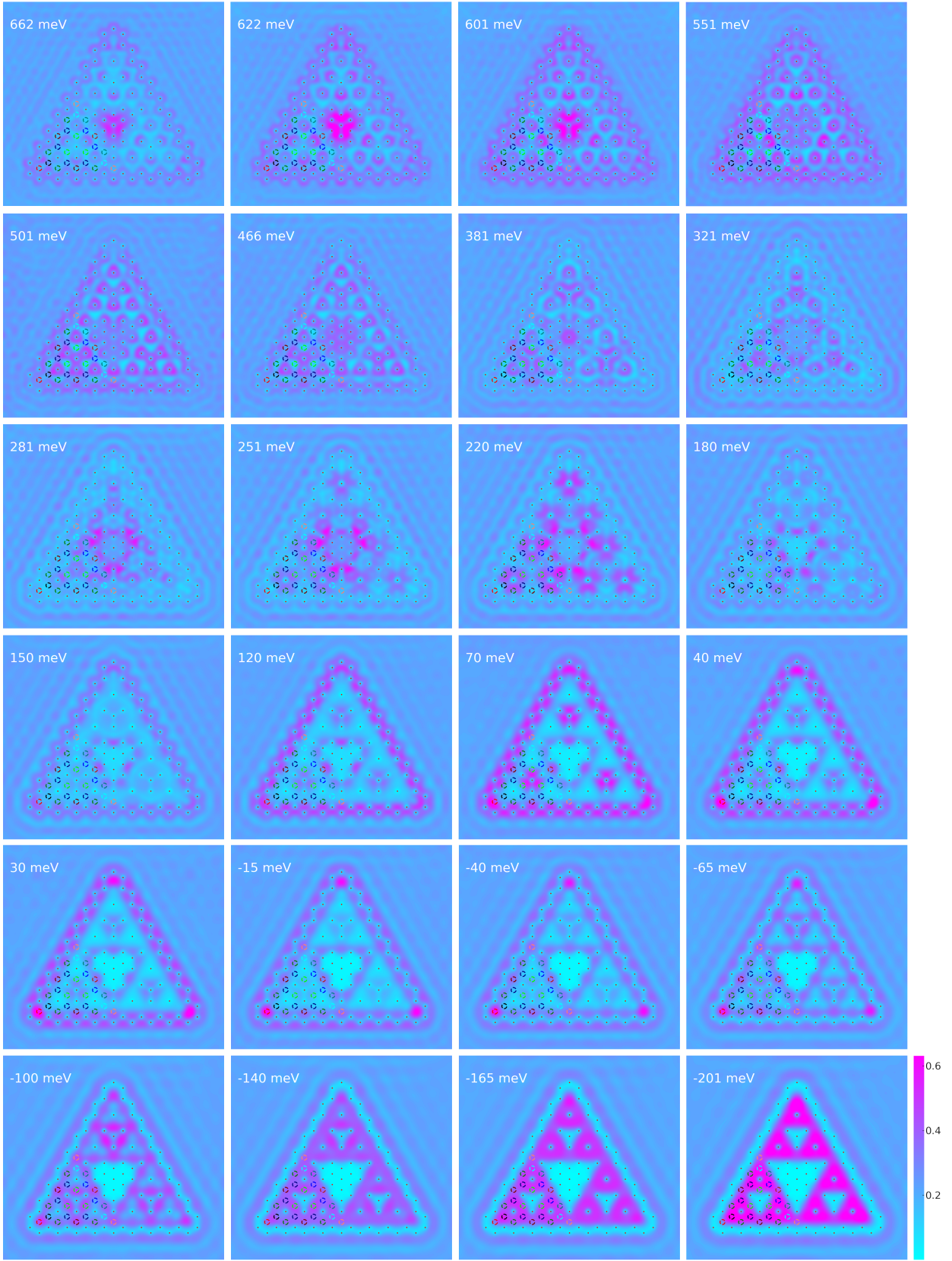

FIG. 16. Theoretical LDOS maps for selected values of energies. The parameters of the simulation are the same as in Fig. 3 of the main text.

### C. Effect of disorder

After describing the general trends of the experimental measurements of the LDOS of Fig. 15, we now take into account the main effects of disorder that we simulated, and connect them to the observations. Firstly, we ran the simulations with a 10% of error for each type of disorder, potential and position, and used a low resolution of 80x80. In this case, the disorder introduced at the LDOS was only visible for the position disorder, while the potential-disorder results remained invariant. Even 5% of error already leads to visible changes for the position disorder, whereas the potential disorder requires an increase of up to 20-25% to start affecting the system (always using the same resolution of 80x80). This suggests that the position disorder is more relevant because small deviations on the position already lead to observable asymmetry on the LDOS. Furthermore, interference patterns are related to the difference in the path performed by the wavefunctions, giving rise to a difference in phase. Thus, the observation of the LDOS asymmetry for small variations of the scatterers positions is a consequence of the interference phenomenon. The free electron gas is scattered back at different positions, generating this asymmetry.

However, when the resolution is increased to 200x200, we see that the potential disorder also generates an asymmetry to the LDOS. Both types of disorder induce similar features; phases that originally had a difference in energy become mixed, and depending on the implementation, they start to form at smaller regions of the Sierpinski fractal. This agrees with the experimental LDOS results. Now, we reinterpret the LDOS maps shown in Fig. 15 in the light of the theoretical results for disorder in the muffin-tin model. We will refer to the three sub-regions corresponding to previous generations by the top, left, and right.

Starting at the lowest value of energy $E = -507$ meV, we observe that the bulk phase is starting to form. The pink phase (high LDOS) tends to occupy the whole region and is intensified at $E = -447$ meV and $E = -427$ meV. These are the values where the pink phase is more homogeneous. Increasing energy, the bulk phase exhibits a lot of asymmetries, which are particularly more present at the left and right triangles, see e.g., $E = -286$ meV, $E = -266$ meV, and $E = -245$ meV. These left and right regions are then the first at which the phase that looks like a 3-clover leaf forms. It is only later, at $E = -165$ meV, that the same phase starts to form at the top triangle. At $E = -125$ meV and $E = -104$ meV, they show more similarity. At $E = -84$ meV the top triangle is intensified (similarly to the left), while the right is beginning to disappear. This could be related to the right corner mode, which is not observed. At $E = -44$ meV, the map shows that the corner mode arises at the top. Then, the left corner starts to be populated from $E = -4$ meV to $E = 16$ meV. At $E = 57$ meV, an edge mode starts to appear at the bottom. From $E = 97$ meV to $E = 137$ meV, this edge mode is also located at the right and left sides of the larger triangle. It forms similarly to the inner edge, which is first populated at the right and top at $E = 198$ meV, and then spreads to left sides at $E = 218$ meV.

### D.  Straight boundaries

Here, we study how the free electron gas responds to another type of confinement. Instead of having scatterers defining an effective potential, we draw continuous contours that delineate the region where the potential is non-zero. Figure 17 shows the potential landscape for the second generation $G = 2$ of a Sierpinski gasket at the yellow region, the potential is $u = 0.9$ eV, while at the black region it vanishes. The colored circles labels the different electronic sites, as before. The small red dots are left to compare the shape with the previous positions of the scatterers. In addition, we have rotated the triangle by 10°with respect to the boundaries, to investigate another type o disorder.

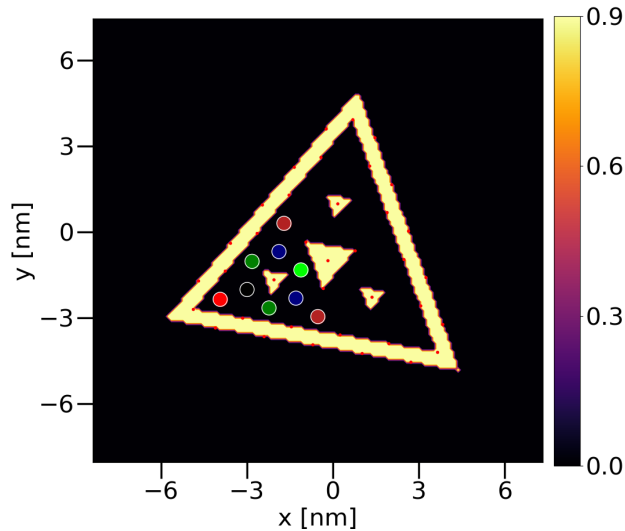

FIG. 17. Sketch of an abrupt potential following the triangular straight wall confinement. The number of grid points in both axes are $n_x = n_y = 150$, the width of the potential wall is $d = 0.62$, the rotation angle is $\theta = 10°$, and the potential height $u = 0.9$ eV.

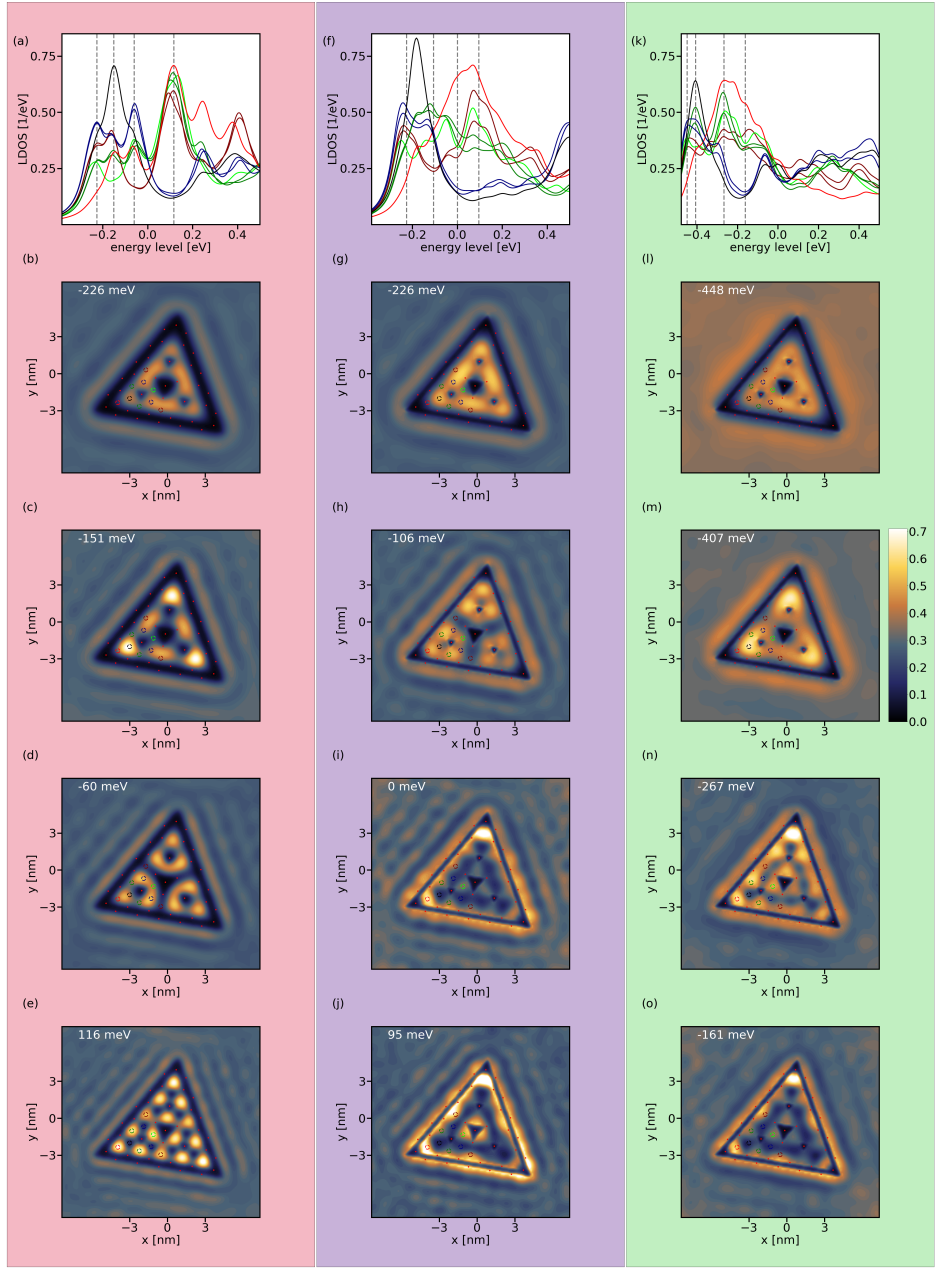

FIG. 18. Theoretical LDOS maps for the second generation (G=2) Sierpinski triangle using a straight boundary confinement. The calculations shown in the first column (pink) include only the kinetic term, while the second column (purple) also include the ISOC, and the third (green) include all three terms, kinetic, intrinsic, and Rashba SOC. The number of grid points in both axes is $n_x = n_y = 150$, the number of waves 750, the width of the wall is $d = 0.62$ nm, the rotation angle is $\theta = 10°$, the onset energy $u_s = 388$ meV, and a potential height $u = 0.9$ eV. The electron effective mass $m_{\text{eff}} = 0.42$, and the intrinsic and Rashba SOC are $\lambda_{\text{ISOC}} = 10^6$ and $\lambda_{\text{Rsh}} = 10^9$, respectively.

Figure 18 presents the LDOS calculated including only the kinetic term (pink background), kinetic and ISOC (purple), and all of them plus Rashba (green). Comparing with Fig. 2 of the main text, we see that the rotation of the lattice destroys the $C_3$ symmetry of the Sierpinski gasket for some phases. This is the case for the bulk phases visible in Figs. 18(g-h) and (l-m). More importantly, the rotation creates an asymmetry between the corner states, seen in Fig. 18(i). Further, the topological phases in this geometry seem to be more resilient against the Rashba SOC. Using the value of $\lambda_{\text{Rsh}} = 10^9$, we see [Fig. 18(l-o)] features that resemble the ones for the system with only ISOC. In opposition, the same value of $\lambda_{\text{Rsh}} = 10^9$ is already enough to destroy the features introduced by the ISOC for the muffin-tin method (see Fig. 2 of the main text). Notice that the onset energy $u_s = 388$ meV is slightly different than the one used for the Gaussian potential.

## IX. BROADENING

The experimental peaks in the LDOS are sharper than the theoretical ones. Using a smaller value for the broadening in the simulations, one obtains a LDOS spectrum more similar to the experimental one, but oscillating peaks arise as an artifact. Figure 19(a)-(d) depict the LDOS curves for four different values of broadening $b = \{0.04, 0.02, 0.01, 0.005\}$ eV, respectively. When decreasing the value of $b$, the read peak at zero energy splits in three different peaks, with a width and heights following better the trend of the experimental curves. However, they do not capture completely the measured V shape of the corner modes shown in Fig. 1 of the main paper. To obtain this feature, we would need to consider an even smaller broadening, but then the FWHM decreases and more spurious oscillations appear.

We also noticed that the curves change, depending on the underlying grid. The results shown at Fig. 19(a)-(d) correspond to $n_x = n_y = 120$. In Figure 19(e)-(h) we exhibit the effect of changing the number of points $n_x = n_y = \{80, 120, 150, 200\}$, but keeping $b = 0.01$ eV. With more grid points the three read peaks shown in Fig. 19(b) decouple and start splitting further. Although decreasing the broadening and the grid yields a better agreement with the LDOS curves, when plotting the 2D maps, the features are not well discernible. Several phases become less pronounced, and they start to appear at sub-regions of the Sierpinski triangle for different energy scales. For this reason, we concluded that a broadening $b = 0.04$ was the best compromise to capture all phases observed experimentally.

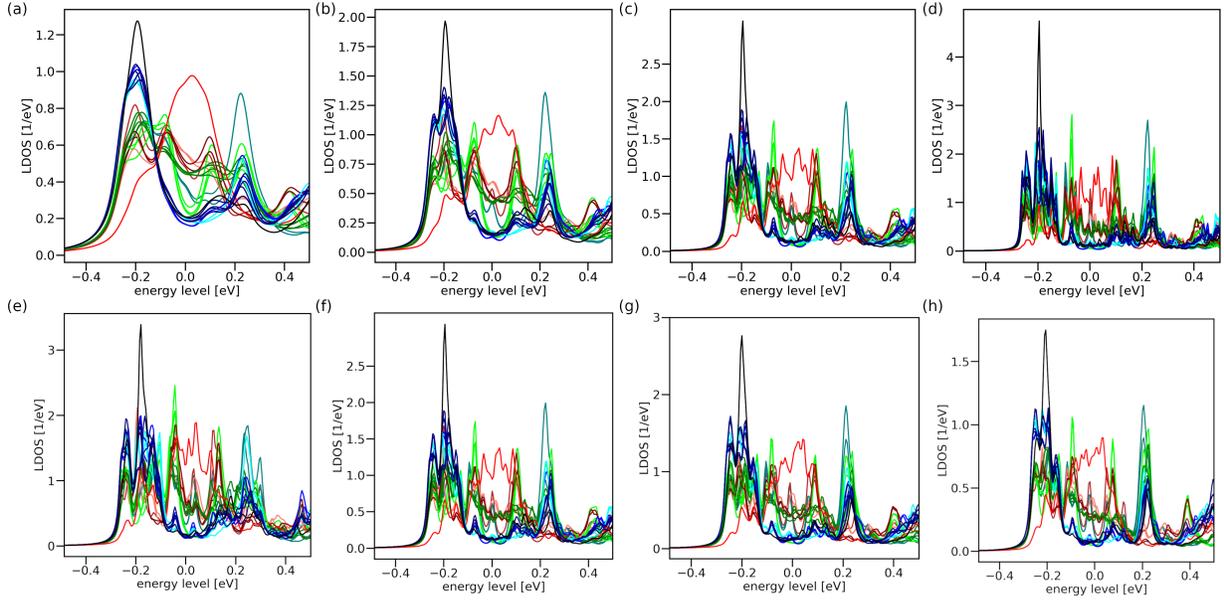

FIG. 19. LDOS curves with (a)-(d) different broadening $b = \{0.04, 0.02, 0.01, 0.005\}$ eV and fixed lattice grid $n_x = n_y = 120$; (e)-(d) fixed broadening $b = 0.01$ eV and different lattice grid $n_x = n_y = \{80, 120, 150, 200\}$, respectively.

## X. TIGHT-BINDING

In the main text, we show scaled tight-binding LDOS to portray data in a consistent manner to both, experiments and muffin-tin results. However, if we allow the scaling per plot to be free, the features observed in the experiments are more evidently visualized in the tight-binding results, as shown in Fig. 20. Note that although the LDOS maps are similar to those observed in the experiments, they are not exactly the same. For example, in the tight-binding LDOS maps, it is not possible to resolve an isolated outer edge mode; there are always some inner edge modes present concomitantly. This is different from the muffin-tin calculations and from the experimental results. On the other hand, the "blobby' phase is better captured within the tight-binding description.

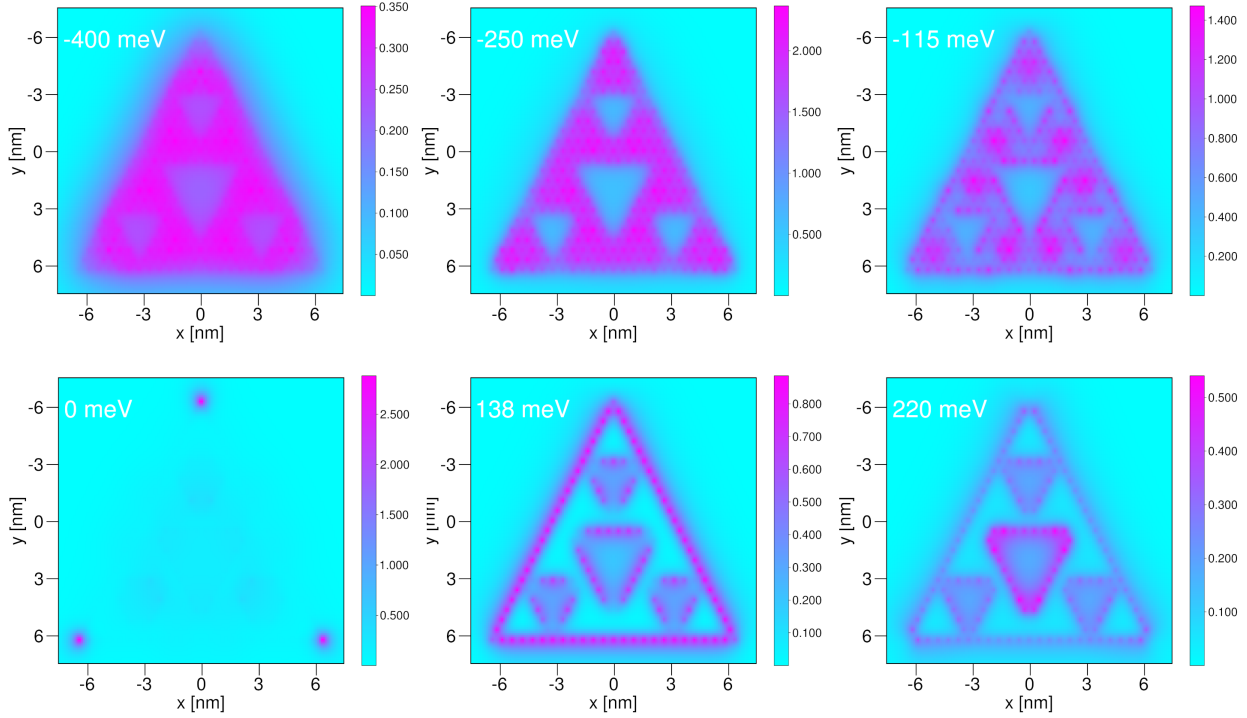

FIG. 20. Free-scale version of the tight-binding LDOS plots shown in Fig. 3 of the main text.

---